\documentclass[
aps,
pra,
longbibliography,
superscriptaddress,
 amsmath,amssymb,
twocolumn,
]{revtex4-1}

\usepackage{graphicx}
\usepackage{dcolumn}
\usepackage{bm}
\usepackage{upgreek}
\usepackage{color}
\usepackage{hyperref}
\DeclareMathAlphabet{\mathpzc}{OT1}{pzc}{m}{it}
\usepackage{enumitem}   

\newcommand{\RR}{\right}
\newcommand{\LL}{\left}
\newcommand{\m}{\mathrm}
\newcommand{\dg}{\dagger}
\newcommand{\eref}[1]{Eq.~(\ref{#1})}
\newcommand{\fref}[1]{Fig.~\ref{#1}}
\newcommand{\puoli}{\frac{1}{2}}
\newcommand{\rthz}{/\sqrt{\m{Hz}}}

\usepackage[normalem]{ulem}

\newcommand{\MS}[1]{{#1}}
\newcommand{\JM}[1]{{#1}}

\setlist[enumerate]{itemsep=-1mm}

\begin{document}

\title{Prospects for observing gravitational forces between nonclassical mechanical oscillators}

\author{Yulong Liu}
\affiliation{Quantum states of matter, Beijing Academy of Quantum Information Sciences, Beijing 100193, China}
\affiliation{Department of Applied Physics, Aalto University, P.O. Box 15100, FI-00076 AALTO, Finland}

\author{Jay Mummery}
\affiliation{Department of Physics, The University of Western Australia 35 Stirling Highway, Crawley WA 6009, Australia}

\author{Jingwei Zhou}
\affiliation{Department of Applied Physics, Aalto University, P.O. Box 15100, FI-00076 AALTO, Finland}%

\author{Mika A. Sillanp\"a\"a}
 \email{Mika.Sillanpaa@aalto.fi}
\affiliation{Department of Applied Physics, Aalto University, P.O. Box 15100, FI-00076 AALTO, Finland}%


\date{\today}

\begin{abstract}
Interfacing quantum mechanics and gravity is one of the great open questions in natural science. Micromechanical oscillators have been suggested as a plausible platform to carry out these experiments. We present an experimental design aiming at these goals, inspired by Schm\"ole \textit{et al.}, Class.~Quantum Grav.~\textbf{33}, 125031 (2016). Gold spheres weighing on the order a milligram will be positioned on large silicon nitride membranes, which are spaced at submillimeter distances from each other. These mass-loaded membranes are mechanical oscillators that vibrate at $\sim 2$ kHz frequencies in a drum mode. They are operated and measured by coupling to microwave cavities. \MS{First, we show that it is possible to measure the gravitational force between the oscillators at deep cryogenic temperatures, where thermal mechanical noise is strongly suppressed.} We investigate the measurement of gravity when the positions of the gravitating masses exhibit significant quantum fluctuations, including preparation of the massive oscillators in the ground state, or in a squeezed state. We also present a plausible scheme to realize an experiment where the two oscillators are prepared in a two-mode squeezed motional quantum state that exhibits nonlocal quantum correlations and gravity the same time. Although the gravity is classical, the experiment will pave the way for testing true quantum gravity in related experimental arrangements. In a proof-of-principle experiment, we operate a 1.7 mm diameter Si$_3$N$_4$ membrane loaded by a 1.3 mg gold sphere. At 10 mK temperature, we observe the drum mode with a quality factor above half a million at 1.7 kHz, showing strong promise for the experiments. Following implementation of vibration isolation, cryogenic positioning, and phase noise filtering, we foresee that realizing the experiments is in reach by combining known pieces of current technology.
\end{abstract}

\maketitle


\section{Introduction}

It was not until the 1930's that it became legitimate to ask whether quantum-mechanics that holds for elementary particles, could also be used to describe degrees of freedom in larger bodies \cite{Cat}. Such ``secondary" macroscopic quantum phenomena are distinguished from, for example, magnetism or many other properties of solid matter that are of very quantum origin. In contemporary experimental research, secondary macroscopic quantum-mechanical phenomena, for example superpositions, entanglement, energy quantization and zero-point noise, are routinely observed with large molecules \cite{Arndt1999}, atomic ensembles \cite{Polzik2001Ent}, superconducting microcircuits \cite{nakamura}, or in particular, in the motion of micromechanical oscillators, see e.g.~\cite{ClelandMartinis}. As compared to normal atomic quantum phenomena, it is much more challenging to create and observe macroscopic quantum effects since they are more fragile with respect to environmental decoherence.

The theory of general relativity, in the same way as quantum mechanics in its own realm, has been extraordinarily successful in describing huge objects interacting via gravitational force. One of the greatest results predicted by general relativity was the discovery of gravitational waves by the LIGO laser interferometer \cite{LIGO}. Regrettably, general relativity and quantum mechanics do not get along well at all. For example, time in quantum mechanics is an absolute quantity, whereas in relativity it is a dynamic variable. The theory resulting from blindly quantizing general relativity is not renormalizable and hence far from satisfactory at a fundamental level. It is not even known if gravity is a quantum-mechanical entity whatsoever, or whether it is somehow disconnected from the rest of the universe that obeys quantum mechanics. One can say the main point in the studies is to find out if the gravitational field created by a quantum superposition state is a quantum superposition of the two fields.

On the experimental side, the situation is equally bleak. Quantum gravity, if such an entity exists, plays a role at the respective Planck length, time, or energy scales of approximately $10^{-35}$ meters, $10^{-44}$ seconds, and $10^{19}$ GeV. Direct experimental access to these extraordinary scales is not possible in the foreseeable future.

Interestingly, it has been realized recently that indirect measurements could probe the gravity-quantum interface. Oscillating physical bodies in the motional quantum regime  - vaguely speaking with a thermally excited phonon number less than one - might be well suited for the purpose \cite{AspelmeyerPlanck,Marin2013QGR,Marin2016QGR,Rovelli2019,Mann2019QGR,Yang2020qgrav}. Since 2010 \cite{ClelandMartinis}, nonclassical motional states are quite routinely achieved in various types of nano- and micromechanical oscillators in the size range up to approximately 10 microns, weighing up to a nano-gram.  In many experiments, a setup known as cavity optomechanics has been utilized, where mechanical oscillations are coupled to electromagnetic fields located inside cavity resonators of various geometries and frequencies \cite{OptoReview2014}. A benchmark would be to prepare a quantum state in an oscillator whose mass exceeds the mass equivalent $\sim 20$ $\mu$g of the Planck energy, which represents a plausible crossover above which quantum states have been hypothesized to decohere \cite{Penrose,Reynaud2006gravity}. Experimental preparations have been going on to cool oscillators with this heavy effective masses towards the motional ground state \cite{Heidmann2006,LIGO2009kgQM,Santos2017,Matsumoto2019pendulum,Matsumoto2020pendulum,LIGO2020squeeze}. Microwave cavity optomechanics \cite{Lehnert2008Nph} offers another possibility as a side product from the more complex scheme we discuss below.

Letting aside quantum gravity for a moment, measuring the effect of Earth's gravity on a small quantum object is an interesting goal, achieved with neutron interferometry since the 1970's \cite{Overhauser1975,Strelkov2002neutron}, and followed by experiments on cold atoms \cite{Chu1999,Rosi2014atomGR}. The next and more demanding step is arguably verifying the gravitational force between very small objects.
Being able to make such measurement is likely a prerequisite for more delicate measurements that involve both gravity and quantum phenomena inside the system. Since the Cavendish experiment at the end of the 18$^\m{th}$ century, the masses between which gravitational forces have been directly measured, have got smaller \cite{Ritter1990}, reaching sub-gram source masses \cite{Tan2016torsion,Tan2020torsion} at $\sim 200$ $\mu$m distance. Even lower source masses consisting of a 0.2 gram hole have been claimed \cite{Heckel2020}. The aim in these torsion balance experiments and in many earlier \cite{Chiaverini2003GR,Decca2005torsion,Tobar2015GR,Kostelecky2015torsion} has been to test deviations from the Newton's law of gravity at very small distances, although typically they have been null measurements without directly measuring the gravity.


\begin{figure}[h]
  \begin{center}
   {\includegraphics[width=0.95\columnwidth]{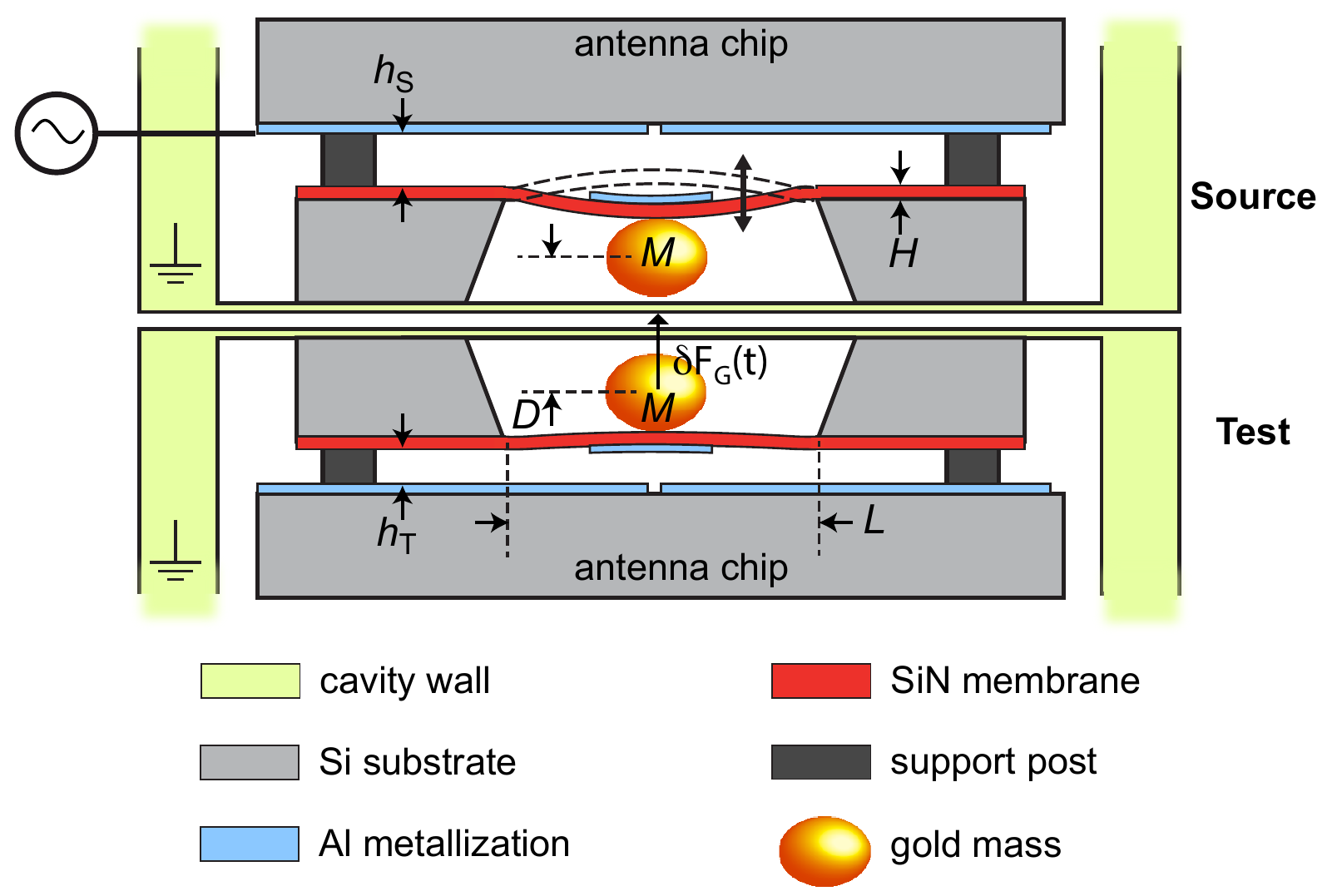} }
    \caption{\textit{Schematics of the proposed experiment.} Two flip-chip assemblies with mass-loaded  high-stress silicon nitride membranes are positioned at submillimeter distance from one another. There can be an appreciable gravitational force between the gold spheres, which can be observed using microwave optomechanical techniques, for which the metallized antennas mediate coupling.}
    \label{fig:scheme1}
 \end{center}
\end{figure}

It has been proposed that micromechanical devices could push the mass limit of detecting a purely Newtonian gravitational force down to the milligram range of source masses \cite{Aspelmeyer2016grav}. Very recently, the scheme was experimentally demonstrated \cite{Aspelmeyer2020gra}. According to the scheme, the source mass is strongly actuated. The time-dependent gravitational pull can be sensed by taking advantage of resonant enhancement, where the test mass is sitting on a vibrating beam. In the current paper we describe a concrete proposal that can realize, first of all, the measurement of self-gravity in a system of mg particles. As compared to Ref.~\cite{Aspelmeyer2016grav}, which considered room-temperature operation and frequencies $< 100$ Hz, our scheme aims on higher frequencies of several kHz and deep cryogenic temperatures in order to reach the quantum regime. 

A  sketch of the device scheme is shown in \fref{fig:scheme1}. We utilize microwave cavity optomechanics that is naturally compatible with cryogenics, which also offers the benefit of reduced dissipation. The main interest is on finding a way to determine gravity when the source mass or the test mass, or both, are localized at the vacuum level and thus exhibit significant vacuum fluctuations. The possible experiment will not yet create superposed gravitational fields because the positions of the masses are not in any appreciably spatially separated superposition. In our view, however, this detailed proposal is unique in the sense that its implementation seems readily in reach to touch the interface between quantum mechanics and gravity, and it would present an important step on this very fundamental line of research.


\section{Theory and modeling}

In this section, we discuss various aspects on the underlying theory, and present results of numerical modeling and compare these against achievable experimental parameters.

A decisive prerequisite for these sensitive measurements will be the elimination of thermal noise of the mechanical mode. In thermal equilibrium, an oscillator of a resonance frequency $\omega_0$ has the thermally excited phonon number $n_m^T = \left[\exp(\hbar \omega_0 /k_B T )-1\right]^{-1}$. Low-order flexural mechanical modes in whatever system tend to be of a much lower frequency than say electromagnetic modes, and therefore usually $n_m^T \gg 1$ even at deep cryogenic temperatures. Thus, reaching the pertaining quantum regime in mechanical systems in general is challenging. However, reaching this limit using various cooling techniques in cavity optomechanics is these days quite standard. The basic requirement is that the phonon number $n_m \ll n_m^T$ associated to a remaining thermal population is on the order one, or better, much less than one.

\subsection{Microwave cavity optomechanics}

Traditionally, cavity optomechanics has studied the interaction between a mechanically vibrating mirror with the light field inside an optical Fabry-Perot resonator. The coupling between light and motion is due to radiation pressure. 
The important parameters include the single-photon radiation-pressure coupling energy $g_0$, the mechanical resonance frequency $\omega_0$, mechanical damping rate $\gamma$, and quality factor $Q$, as well as the resonance frequency $\omega_c$ and damping rate $\kappa$ of the cavity. The Hamiltonian is then
\begin{equation}
H = \omega_{c} a^\dag a+ \omega_0 b^\dag b + g_{0} a^\dag a \left(b+b^\dag \right) \,,
\label{eq:H}
\end{equation}
where the creation and annihilation operators for photons (phonons) are denoted by $a$, $a^\dg$ ($b$, $b^\dg$).

Complicated quantum state control can be achieved via combinations of the basic operations created by applying coherent electromagnetic waves at frequencies corresponding to Stokes or anti-Stokes motional sidebands of the cavity. Although the $g_0$ parameter is small compared to other scales, the pumping tones can also be seen as creating a large effective coupling $\mathcal{G} = g_0 \sqrt{n_P }$  where $n_P \gg 1$ is the pump-induced photon number. One also defines the cooperativity $\mathcal{C}=4G^2/(\kappa \gamma)$ that describes the interaction strength, and a corresponding optical damping $\Gamma_{\m{opt}}=4\mathcal{G}^2/\kappa$. Nearly arbitrary (Gaussian) quantum operations can be performed by suitable pumping with many coherent tones. The scheme is easily extended to multi-partite systems. A carefully designed pumping can also be seen to turn the cavity into a dissipative reservoir such that the oscillator sees a dissipative bath that pulls the oscillator into a nontrivial steady state.

Some of the important experimental advances in the field of quantum cavity optomechanics include sideband cooling to the ground state \cite{Teufel2011b,AspelmeyerCool11}, entanglement between microwave light and motion \cite{LehnertEnta2013}, radiation-pressure shot noise and optomechanical squeezing \cite{RegalShot,Painter2013Sq,Kippenberg2017SquA,SqueezeAmp,Schliesser2018sql}, non-Gaussian mechanical states \cite{Aspelmeyer2017herald}, motional entanglement between mechanical oscillators \cite{Entanglement,Groblacher}, Bell test \cite{Groblacher2018Bell}, and room-temperature quantum behavior \cite{Purdy2017,Kippenberg2017Squ}.

The basic method to cool down mechanical oscillators in cavity optomechanics at large is sideband cooling. The coherent pump tone is applied at a frequency that is one $\omega_0$ lower than the cavity frequency. Then, phonons are preferentially scattered into the cavity. In order for this protocol to work well, the  good-cavity (resolved-sideband) limit $\kappa \ll \omega_0$ has to be satisfied. Including bad-cavity correction (the last term below), the final thermal phonon occupation of the mode is given by
\begin{equation}
n_m = n_m^T \frac{\gamma}{\gamma + \Gamma_{\m{opt}}} + \LL(\frac{\kappa}{4 \omega_0} \RR)^2 \,.
\label{eq:sbcool}
\end{equation}

Microwave-frequency “cavities” are one of the leading platforms to do cavity optomechanics in the wide sense. Usually these have been realized as on-chip superconducting transmission line resonators. An overwhelming benefit of the microwave-frequency realization is its compatibility with standard cryogenic instrumentation. The mechanical resonators in this setup have usually been either nanowires \cite{Lehnert2008Nph,Painter2016SiN,Fink2017circul,Collin2019demag}, or aluminum drumheads \cite{Teufel2011a,Teufel2011b,Schwab2014QND,Bernier2017,Entanglement}. A very promising recent advance has been achieved with high-stress silicon nitride membranes of several hundred microns in diameter that were implemented inside 3D microwave cavities \cite{Steele3D,Nakamura3D}, which is the setup we are interested in.

The analog of radiation-pressure coupling arises through the fact that a conducting mechanical oscillator modulates the cavity capacitance. Usually the capacitor is an approximate plate capacitor defined by a membrane oscillator and a counter electrode that is connected to the cavity. For example in \fref{fig:scheme1}, there are two such movable capacitors, with the vacuum gaps denoted as $h_S$ and  $h_T$.

The single-quantum radiation-pressure coupling, here given in frequency units, becomes
\begin{equation}
\label{eq:g1}
   g_0 \equiv \frac{d \omega_c}{d x} x_{\m{zp}} = \frac{\omega_c}{2C } \LL( \frac{d C}{d x } \RR) x_{\m{zp}} \,,
\end{equation}
where $C$ is the total capacitance of the $LC$ cavity resonator, and $x_{\m{zp}} = \sqrt{\frac{\hbar}{2 M \omega_0}}$ is the mechanical zero-point oscillation amplitude, where $M$ is the effective mass. 

If we suppose the empty cavity's capacitance dominates over the movable one, we can write \eref{eq:g1} as
\begin{equation}
\label{eq:g1b}
   g_0    \simeq \frac{\omega_c}{2C } \LL( \frac{\varepsilon_0 A}{h^2} \RR) x_{\m{zp}} \,,
\end{equation}
where $A$ is the electrodes' surface area, and $h$ is the vacuum gap. This shows that a narrow vacuum gap is a critical parameter for maximizing the coupling.

\subsection{Mass-loaded membrane}

We select stoichiometric, high-stress Si$_3$N$_4$ (thereafter SiN) as the membrane material due to its well-known excellent mechanical properties, and suitability for (microwave) cavity optomechanics \cite{Steele3D,Nakamura3D}. The internal materials loss is shunted by the high prestress, and very high $Q \sim 10^9$ at MHz frequency can be reached even at elevated temperatures \cite{Schiesser2017,Kippenberg2018dissip,Schliesser2018sql}. Let us consider a rectangular SiN membrane with side length $L$, thickness $H$, density $\rho \simeq 3.2 \cdot 10^3$ kg/m$^3$, and prestress $\sigma \simeq 900$ MPa. Without any mass loading, the lowest drum mode has the frequency $\omega_0/2\pi \simeq \frac{0.71}{L}  \sqrt{\frac{\sigma}{\rho}}$. For example, with $L = 5$ mm, $\omega_0/2\pi \simeq 76$ kHz.

Let us now add the mass loading. A natural material choice is gold due to its high density and well characterized properties. A spherical mass $M$ with radius $R$ is glued in the center of the membrane. The glue is modeled as a short cylinder of radius $r_{\m{glue}} < R$. Due to the added mass overwhelming the membrane mass, the mode frequency will substantially decrease. We are not aware of expression for the frequency of a loaded rectangular membrane. We adapt the corresponding result for the circular eardrum \cite{Fletcher}:
\begin{equation}
\label{eq:f0load}
f_0 = \omega_0/2\pi \approx \sqrt{\frac{\sigma H}{2\pi M \log(L/2r_{\m{glue}})}} \,.
\end{equation}
The frequency depends now on the membrane thickness, but not directly on its width. However, the frequency depends significantly on the participation ratio of the glue spot, and goes down when the spot size decreases. The latter is because the less contact area there is between the mass and the membrane, the less tensile force will pull the mass.

\begin{figure}[h]
  \begin{center}
   {\includegraphics[width=0.99\columnwidth]{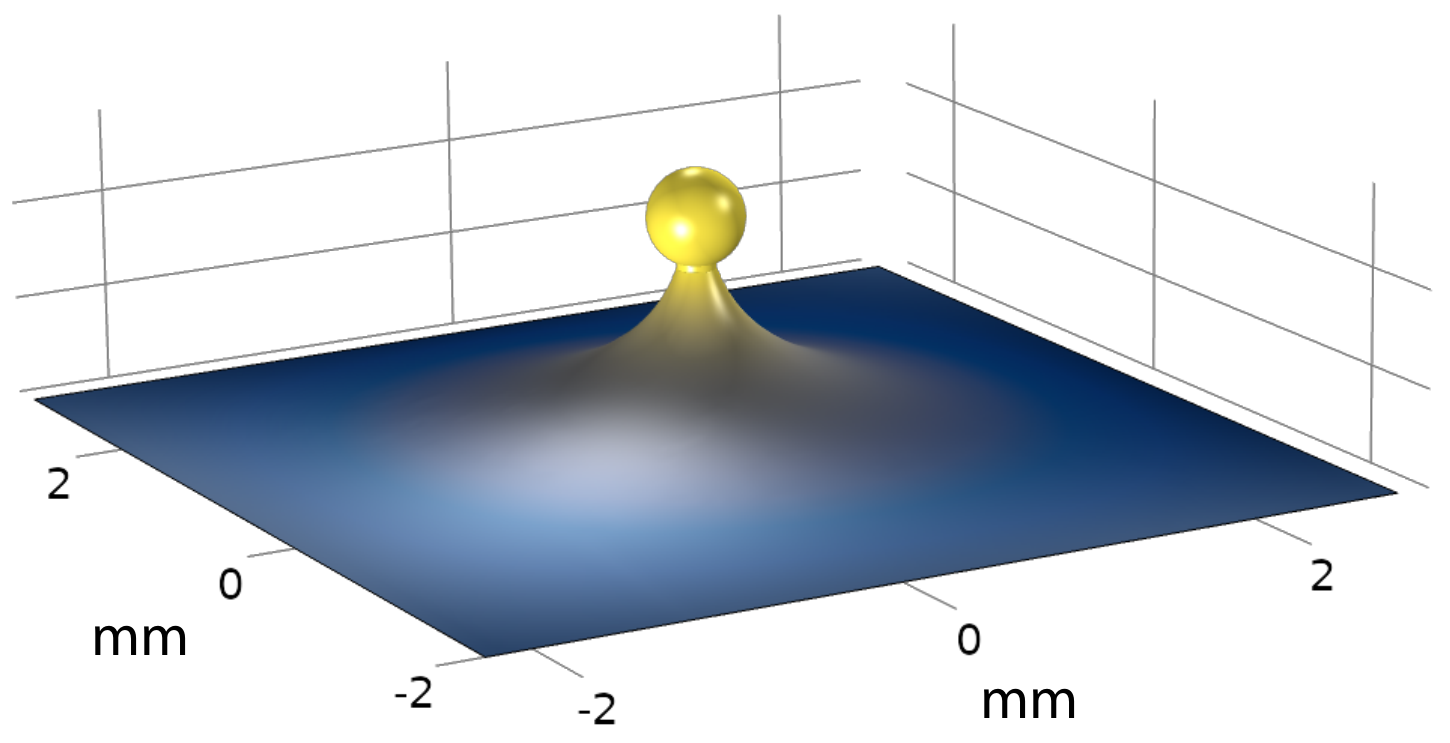} }
    \caption{\textit{Mass-loaded drum mode.} The finite-element simulation shows the profile of the lowest drum mode of an $L = 5$ mm wide, $H = 100$ nm thick, rectangular silicon nitride membrane with a $R = 0.25$ mm radius gold sphere glued in the center. The glue spot radius is $r_{\m{glue}} = 100$ $\mu$m. The mode frequency becomes $f_0 \simeq 1.9$ kHz. 
    }
    \label{fig:sim5mm}
 \end{center}
\end{figure}

We ran finite-element model (FEM) simulations to verify the mode profiles. For the simulation, the glue is modeled as a cylinder connecting the membrane and the sphere. An example mode shape is shown in \fref{fig:sim5mm}. In a couple of tests using different parameter values,  \eref{eq:f0load} holds to within 10 \%. In addition to the main flexural mode, there are two ideally degenerate modes where the sphere wobbles horizontally. Their frequency depends sensitively of the glue spot size, and is typically lower than that of the main mode. There are no other modes within a decade in frequency of the main mode. One can easily verify from the simulations that the weight of the sphere does not significantly deflect the membrane in comparison to a typical vacuum gap, neither does it affect the mode profiles almost independently on the orientation of the chip with respect to horizontal. These are due to the high tension in the membrane.

With the high mass load, the optomechanical coupling, \eref{eq:g1} will be reduced because the zero-point motion will go down, while the electrode geometry can be assumed unchanged. The dependence of $g_0$ on the mass load is only weak, $g_0 \propto M^{-1/4}$. Using the discussed parameters, $g_0$ amounts to about 30 \% of that of a bare membrane. With a 0.5 $\mu$m vacuum gap between the electrodes, we have on the order $g_0/2\pi \sim 0.2$ Hz that is still a very decent value as will be discussed below in light of other experimental parameters.

A critical consideration is how the mechanical losses are influenced by the exceptional geometry. It can be expected that especially the glue, but also gold, can have high losses even at cryogenic temperatures. As seen in \fref{fig:sim5mm}, the mode profile exhibits a sharp kink at the joint between the membrane and the glue. A large curvature is known to enhance the losses of the mode \cite{Kotthaus2010loss,Schiesser2017}. In the FEM simulation we take care that the mesh around the kinks is small ($0.5$ $\mu$m) in comparison to the resulting bending radius ($\sim 3$ $\mu$m). For SiN, we use a low-temperature loss factor $\sim 10
^{-5}$. Using the parameters mentioned in \fref{fig:sim5mm}, we obtain that SiN losses set $Q \lesssim 2\times 10^{8}$. Although curvature is strongly enhanced at the kink, it is the same time decreased in the regions away from the center. This could also reduce the losses due to clamping, although this was not considered in the simulations.

For gold or epoxy, the low-temperature loss is not readily available. For the epoxy, we use that of PMMA, on the order $\sim 10
^{-3}$ \cite{Thompson2002materials}, and assume the same for gold. Including all the estimated losses, the total $Q$ value is obtained from $Q^{-1} = \sum Q_i^{-1}$ summing all the considered loss channels $i$. In the current example, $Q \sim 7 \times 10^{6}$ for the main mode, strongly dominated by losses of the glue. The contributions by different loss channels are listed in Table \ref{tab:simQ}, for a couple of plausible geometries. It is seen that a large glue radius and thin membrane are beneficial, and that gold losses are nearly irrelevant. The wobbling modes are affected by nearly an order of magnitude more by all the losses, and in the end they have $Q \sim  10^{6}$.

The high predicted $Q$ values, acquired without exhaustive optimization of the geometry, show a great promise for the experimental realization. Below in the discussion, we suppose $Q = 10^7$ that seems achievable in real life. By optimizing the contact between the membrane and the mass, one could further achieve an order of magnitude increase in $Q$.


\begin{table}[ht]
\caption{\emph{Mechanical $Q$ values due to materials losses}. The last column gives the total $Q$ with the parameters $\eta_{\m{SiN}} = 10^{-5}$, $\eta_{\m{Au}} = 10^{-3}$, $\eta_{\m{glue}} = 10^{-3}$.}
\begin{center}
\begin{tabular}{|c|c|c|c||c|c|c|c|}
\hline\hline
 $L$   &   $H$ &   $r_{\m{glue}}$  &$f_0$ & $Q_{\m{SiN}}$   &   $Q_{\m{Au}}$   &   $Q_{\m{glue}}$  & $Q$   \\
  (mm)   &   (nm)&   ($\mu$m)  & (kHz)     &  &    &    &   \\   \hline
 5    &   50 &   100 &  1.3    & $\frac{3 \times 10^{3}}{\eta_{\m{SiN}}}$   &   $\frac{2.3 \times 10^{5}}{\eta_{\m{Au}}}$   &   $\frac{2 \times 10^{4}}{\eta_{\m{glue}}}$ & $17 \times 10^{6}$  \\ \hline
  5    &   100 &   50 &   1.7    & $\frac{8 \times 10^{2}}{\eta_{\m{SiN}}}$   &   $\frac{6 \times 10^{4}}{\eta_{\m{Au}}}$   &   $\frac{5 \times 10^{3}}{\eta_{\m{glue}}}$ & $4 \times 10^{6}$  \\ \hline
5    &   100 &   100 &  1.9    & $\frac{1.4 \times 10^{3}}{\eta_{\m{SiN}}}$   &   $\frac{1.1 \times 10^{5}}{\eta_{\m{Au}}}$   &   $\frac{8 \times 10^{4}}{\eta_{\m{glue}}}$ & $7 \times 10^{6}$  \\ \hline
 5    &   100 &   200 &   2.1    & $\frac{2 \times 10^{3}}{\eta_{\m{SiN}}}$   &   $\frac{3 \times 10^{5}}{\eta_{\m{Au}}}$   &   $\frac{1.1 \times 10^{4}}{\eta_{\m{glue}}}$ & $11 \times 10^{6}$  \\ \hline
  5    &   200 &   200 &   2.9    & $\frac{1.1 \times 10^{3}}{\eta_{\m{SiN}}}$   &   $\frac{1.6 \times 10^{5}}{\eta_{\m{Au}}}$   &   $\frac{5 \times 10^{3}}{\eta_{\m{glue}}}$ & $5 \times 10^{6}$  \\ \hline
  1.7    &   50 &   100 &   1.60    & $\frac{1.6 \times 10^{3}}{\eta_{\m{SiN}}}$   &   $\frac{1.6 \times 10^{5}}{\eta_{\m{Au}}}$   &   $\frac{1.3 \times 10^{4}}{\eta_{\m{glue}}}$ & $11 \times 10^{6}$  \\ \hline
\end{tabular}
\label{tab:simQ}
\end{center}
\end{table}

\subsection{Displacement detection}

Let us recap the results for the continuous position monitoring of a single oscillator. Such a measurement is typically carried out by probing a cavity-optomechanical system at the resonant frequency of the cavity. The aim is to find out how the different known noise contributions in the gravitational force measurements will affect the sensitivity. In the deep quantum case, it turns out the measurement quantum backaction will pose limitations for the sensitivity. It could then be beneficial to use quantum backaction evading (BAE) measurement strategies \cite{Caves1980,Onafrio1996force}.

We write the linearized optomechanical Hamiltonian from \eref{eq:H} in terms of position $x$ and momentum $p$ of the oscillator, and the dimensionless quadratures $x_c$, $y_c$ of the cavity:
\begin{equation}
\begin{split}
\label{eq:hquad}
H =  \frac{p^2}{2M} + \puoli M \omega_0^2 x^2 + \frac{\hbar \omega_c}{4}\left(x_c^2 + y_c^2 \right) + \frac{2 \hbar G}{x_{\m{zp}}} x x_c \,.
\end{split}
 \end{equation}
In addition, there is a noise force $f(t) = f^L(t) + f^E(t)$ driving the oscillator consisting of the thermal plus quantum Langevin force $f^L(t)$,
and of an external noise $f^E(t)$ due to environmental vibrations. The external noise $f^E(t) = M \ddot{x}^E(t)$ can be modeled as an inertial force arising from vibrations of the chip, $x^E(t)$.  Writing this way allows  access to the most concrete technical noise, namely vibration noise in the refrigerator.  The Langevin force has the spectral density
\begin{equation}
\label{eq:fTspectrum}
\begin{split}
S_f^L(\omega) = 
2 \gamma M \hbar \omega_0 \left(n_m^T + \puoli \right) \equiv S_f^T + S_f^{\m{zp}}\,,
\end{split}
\end{equation}
and the spectral density of the total force is $S_f(\omega) = S_f^L(\omega) + S^E_f(\omega)$. Above, the vacuum, thermal, and external noises are
\begin{subequations}
\begin{alignat}{3}
& S_f^{\m{zp}}(\omega) = \gamma M \hbar \omega_0 \label{eq:sfefflist1} \\
& S_f^T(\omega) = 2 \gamma M \hbar \omega_0 n_m^T \label{eq:sfefflist2} \\
& S_f^E(\omega) = M^2 \omega_0^4 S^E_x(\omega) \label{eq:sfefflist3} \,,
\end{alignat}
\end{subequations}
%
and $S^E_x(\omega)$ is the spectral density of cryostat vibrations.

We use the standard input-output modeling of the optomechanical cavity, with the equations of motion:
\begin{equation}\label{eqmot}
\begin{split}
 \dot{x_c} & = -\frac{\kappa}{2} + \sqrt{\kappa} x_\m{c,in}  \\
\dot{y_c} &= - \frac{2 G}{x_{\m{zp}}} x -\frac{\kappa}{2}
+ \sqrt{\kappa} y_\m{c,in} \\
\dot{x} &= \frac{p}{m}  \\
\dot{p} & = - m \omega_0^2 x - \frac{2 \hbar G}{x_{\m{zp}}} x_c - \gamma p + f(t)
\end{split}
\end{equation}
We suppose the cavity mode to be at zero temperature, and the cavity input noises $x_\m{c,in}, y_\m{c,in}$ have the spectrum
\begin{equation}
S_\m{x,in}(\omega) = S_\m{y,in}(\omega) = \puoli \,.
\end{equation}
We further suppose the bad-cavity limit $\kappa \gg \omega_0$ which is also relevant for the experiment, and a fully overcoupled cavity. In this scheme, the cavity $y_c$ quadrature measures the oscillator position $x$, and the measurement backaction affects $p$ directly, and also $x$ via their coupling. All the forces enter via the mechanical susceptibility
\begin{equation}
\label{eq:chi}
\begin{split}
\chi_m(\omega) = & \frac{1}{M \left(\omega_0^2  - \omega^2 - i \gamma \omega\right)} \,,
\end{split}
\end{equation}
that is, their effect is the strongest on resonance. The spectral density of the position fluctuations becomes
\begin{equation}
\begin{split}
\label{eq:Sxlist}
S_x(\omega) & =   |\chi_m|^2 S_f(\omega) + |\chi_{\m{qba}}|^2 S_\m{x,in}(\omega) \\
& \equiv S_x^{\m{zp}} + S_x^T + S_x^E + S_x^{\m{qba}} \,.
\end{split}
\end{equation}
In \eref{eq:Sxlist}, the result is sorted into contributions by the zero-point mechanical noise $S_x^{\m{zp}}$, thermal noise $S_x^T$, vibration noise $S_x^E$, and the backaction noise $S_x^{\m{qba}}$:
\begin{subequations}
\begin{alignat}{4}
& S_x^{\m{zp}} =   \gamma M \hbar \omega_0 |\chi_m|^2  \label{eq:sxlistzp}\\
& S_x^T = 2 \gamma M \hbar \omega_0 n_m^T  |\chi_m|^2 \label{eq:sxlistT}\\
& S_x^E = S_f^E(\omega) |\chi_m|^2 \label{eq:sxlistE}\\
& S_x^{\m{qba}} = \puoli |\chi_{\m{qba}}|^2 \label{eq:sxlistqba} \,.
\end{alignat}
\end{subequations}
The quantum backaction susceptibility grows with the measurement strength (cooperativity):
\begin{equation}
\chi_{\m{qba}} = 2  \sqrt{\mathcal{C} \gamma} \hbar \chi_m \,.
\end{equation}
We now turn on discussing how the mechanical noise spectra appear as an apparent measurement sensitivity after detection. The output field that leaks out from the cavity replicates the mechanical oscillator position. It is given by
\begin{equation}
\begin{split}
& y_{\m{c,out}}(t) =  \sqrt{\kappa} y_c(t) -  y_{\m{c,in}} (t)
= y_{\m{c,in}} (t)- \frac{2 \sqrt{\mathcal{C} \gamma}  }{x_{\m{zp}}  }x(t) \,.
\end{split}
\end{equation}
Additionally, there will be a contribution by the signal lost on the way, and in particular the noise added by the microwave amplifier. These contributions will be described by the added noise $S_{\m{add}}$ in units of quanta. The amplified and detected spectrum becomes
\begin{equation}
\label{eq:SyoutDet}
\begin{split}
S_{\m{y,out}} (\omega) 
& = \puoli + \frac{4 \mathcal{C} \gamma}{x_{\m{zp}}^2} S_x(\omega)  + S_{\m{add}} \,.
\end{split}
\end{equation}
%
We can rearrange \eref{eq:SyoutDet} in order to secure the inferred force noise spectrum,
\begin{equation}
\label{eq:Sfeff}
\begin{split}
S_f^{\m{eff}}(\omega) & \equiv  
\frac{x_{\m{zp}}^2}{4C \gamma   |\chi_m|^2} S_{\m{y,out}} (\omega)  \\
& = S_f^{\m{zp}} + S_f^T + S_f^{E}+ S_f^{\m{qba}} +  S_f^{\m{imp}}  \,.
\end{split}
\end{equation}
%
Here, the backaction noise $S_f^{\m{qba}}$ and the imprecision noise $S_f^{\m{imp}}$ read
\begin{subequations}
\begin{alignat}{2}
& S_f^{\m{qba}} = \frac{2\mathcal{C}}{x_{\m{zp}}^2} \gamma \hbar^2 \label{eq:sfefflist4} \\
& S_f^{\m{imp}} = \frac{\puoli + S_{\m{add}}}{ \frac{4\mathcal{C}}{x_{\m{zp}}^2} \gamma   |\chi_m|^2} \label{eq:sfefflist5} \,.
\end{alignat}
\end{subequations}

The standard quantum limit (SQL) indicates the minimum noise that results from the competition between improving imprecision and increasing backaction while cooperativity becomes larger. Zero temperature is assumed, as well as a noiseless single-quadrature readout of the output field. The latter implies $S_{\m{add}} = 0$. The frequency-dependent, and resonant, SQL are
\begin{equation}
\begin{split}
S_f^{\m{sql}}(\omega)  &= \gamma M \hbar \omega_0 +\frac{\hbar}{ |\chi_m(\omega)|} \\
S_f^{\m{sql}}(\omega_0)  &= 2 \gamma M \hbar \omega_0 \,,
 \end{split}
\end{equation}
respectively. The latter tells that the added noise equals at best the vacuum noise. Using backaction evading (BAE) measurement, the backaction contribution in \eref{eq:Sfeff} is eliminated, and the SQL is no longer a limitation.

\subsection{Observing gravitational force between two oscillators}
\label{sec:observegrav}

Gravitational force is often negligible in physical or chemical processes at the sub-mm scale, not to mention gravitational interaction between particles under that regime. The latter is not surprising given the fact that the electrostatic force between e.g.~a proton and electron is around forty orders of magnitude larger than the gravitational force between them. These numbers seem to indicate that it is utterly impossible to observe gravitational interactions between small particles or even everyday objects, since gravity appears to be totally masked by much stronger electromagnetic forces. This, however, is actually not the case, as hinted already by the Cavendish experiment and later work  \cite{Ritter1990,Tan2016torsion,Tan2020torsion,Heckel2020}. The reason clearly is that electromagnetic fields tend to become confined inside atoms or molecules, or canceled in conducting systems, but gravity cannot be blocked.

According to Newton, the gravitational force between two equal masses $M$ with center-of-mass (COM) distance $D$ is
\begin{equation}
\begin{split}
F_G = \frac{G M^2}{D^2} \,.
\end{split}
\end{equation}
Here, $G \simeq 6.67 \times 10^{-11}$ N (m/kg)$^2$. For example, the gravitational force is $F_G \sim 0.1$ fN between gold spheres weighing $M \sim 1$ mg and separated by $D \sim 1$ mm. Interestingly, this magnitude is well above a detection threshold given the extreme force sensitivities obtained with nanomechanical mass sensors \cite{Bachtold2013}. Although the latter are not relevant for the present problem, the benchmarking gives an indication that measuring such tiny forces is in principle possible. As a static force, however, one cannot distinguish it from other forces, but a suitable modulation that uses resonant enhancement of a test mass motion could accomplish the measurements. Such a proposal was presented in Ref.~\cite{Aspelmeyer2016grav}. In our work we translate the scheme into cryogenic microwave optomechanics, and first analyze the prospects of seeing the self-gravity with mg masses, also considering the quantum limits of detection. 

Recall the scheme in \fref{fig:scheme1}. Both membranes mass-loaded by gold spheres have the same frequency $\omega_0$. Let us call one of the masses the source mass (S). Its motion is actuated sinusoidally by an amplitude $d x_S$, and near-resonantly  (frequency $\omega_G$). The test mass (T) will then feel the time-dependent gravitational force $d F_{G}$ with
\begin{subequations}
\begin{alignat}{2}
d F_G(t)  = & - |d F_{G}| \sin(\omega_G t) \,,\label{eq:dfg1} \\
 |d F_{G}| & = \frac{2 G M^2}{D^3} d x_S \label{eq:dfg2}
\end{alignat}
\end{subequations}
and respond at the motion amplitude
\begin{equation}
\label{eq:dxT}
d x_T = \frac{2 G M}{\gamma \omega_0 D^3} d x_S \,.
\end{equation}
For example, at $Q = 10^7$, $d x_S = 1 \, \mu$m, we have $d x_T \simeq 35$ fm that is nearly three orders of magnitude larger than the zero-point amplitude. Much better imprecision have been reached with optics \cite{Kippenberg2015FB,Schliesser2018FB}, and also with microwaves \cite{Teufel2016ShotN}, so the measurement sounds plausible, although it will be heavily challenged by the mechanical thermal noise as discussed next.

\begin{figure}[t]
  \begin{center}
   {\includegraphics[width=0.9\columnwidth]{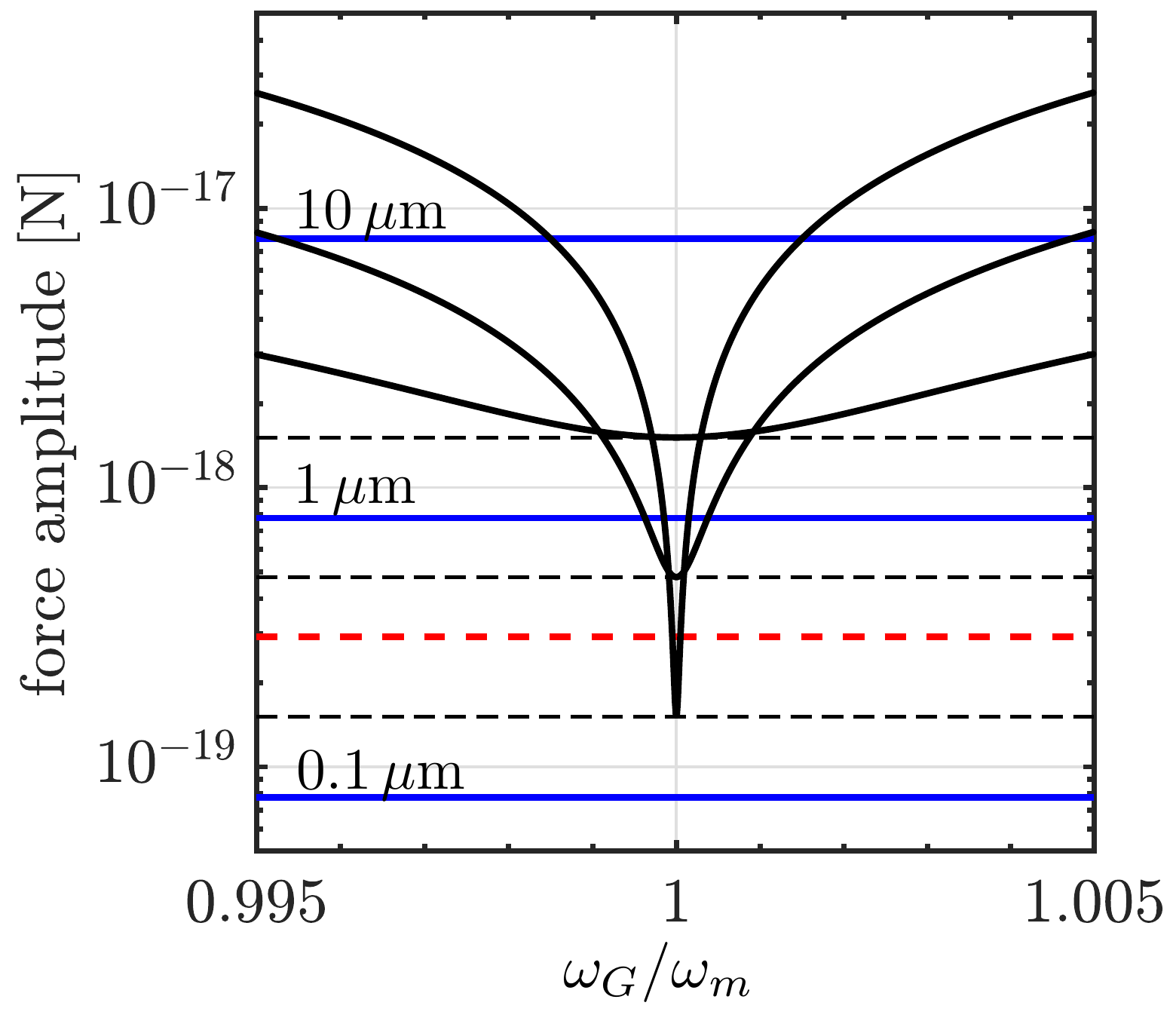} }
    \caption{\textit{Sensitivity to gravitational actuation.} The black curves illustrate the effective force noise, \eref{eq:Sfeff}, at three different mechanical $Q$ values ($Q= 10^6, \, 10^7 \,  10^8$, from top to bottom at center). The dashed horizontal black lines are the corresponding thermal plus vacuum spectra, \eref{eq:fTspectrum}, which meet the corresponding peak on resonance. The red horizontal dashed line is the contribution by external vibration noise with $\sqrt{S_x^E} = 10^{-19}\rthz$. The solid horizontal blue lines are the gravitational signal $S_G$ from the displacements $d x_S$ as labeled. The added noise is $S_{\m{add}} = 10$ quanta. Cooperativity $\mathcal{C} = 10^2$, and integration time is $\tau = 1$ hour.}
    \label{fig:forcenoiseT}
 \end{center}
\end{figure}

\begin{table}[ht]
\caption{\emph{Plausible experimental parameters.} The table lists the parameters (refer to \fref{fig:scheme1}) used in estimating the prospects, unless other parameters are provided in the discussion. It is likely that the mentioned $g_0$ values can be increased by optimizing the cavity geometry.}
\begin{center}
\begin{tabular}{|c|c|c|}
\hline\hline
\emph{symbol} & \emph{description} & \emph{value} \\ 
\hline
$L$ & membrane width & 5 mm \\
\hline
$H$ & membrane thickness & 100 nm \\
\hline
$R$ & gold sphere radius & 250 $\mu$m \\
\hline
$M$ & gold sphere mass & 1.3 mg \\
\hline
$\omega_0$ & mechanical resonance frequency & $2\pi \times 1.9$ kHz  \\
\hline
$x_{\m{zp}}$ & mechanical zero-point amplitude & 60 am \\
\hline
$D$ & center-of-mass distance of spheres & 700 $\mu$m \\
\hline
$h_S$ &  vacuum gap at source oscillator & 2 $\mu$m \\
\hline
$h_S$ &  vacuum gap at test oscillator & 0.5 $\mu$m \\
\hline
$A$ &  electrode surface area &  $(300 \, \mu \m{m})^2$ \\
\hline
$g_{0,S}$ & $g_0$ of source oscillator & $2\pi \times 40$ mHz \\
\hline
$g_{0,T}$ & $g_0$ of test oscillator & $2\pi \times 0.17$ Hz \\
\hline
$\omega_c$ & microwave cavity frequency & $2\pi \times 7$ GHz \\
\hline
$T$ & equilibrium mode temperature & 20 mK \\
\hline
$n_m^T$ & equilibrium phonon number & $2.2 \times 10^5$ \\
\hline
$Q$ & mechanical $Q$ value, both oscillators & $10^7$ \\
\hline
$\Gamma$ & mechanical decoherence rate & $2\pi \times 40$ Hz \\
\hline
\end{tabular}
\label{tab:pars}
\end{center}
\end{table}

The spectral density of the sinusoidal force in \eref{eq:dfg1} is $S_G(\omega) = \frac{\pi}{2} |d F_G|^2  \LL[ \delta(\omega+ \omega_G )+\delta (\omega- \omega_G ) \RR]$. In the measurement this is integrated over frequency, yielding the signal power $P_G = \frac{\pi}{2} |d F_G|^2$. This should be compared to the effective force noise power, \eref{eq:Sfeff}, when recorded within a bandwidth $B$:
\begin{equation}
\label{eq:noiseband}
\begin{split}
P_F = S_f^{\m{eff}}(\omega_G) B \simeq S_f^{\m{eff}}(\omega_G) \tau^{-1} \,,
\end{split}
\end{equation}
where $\tau ~ \simeq B^{-1}$ is the corresponding measurement time. Equation \ref{eq:noiseband} gives the average value of noise power. This is not directly, however, the relevant figure of merit determining the possibility to see a signal peak in the detected spectrum, but instead the peak should be compared to the ``noise of noise" $\sigma_F$ viz.~standard deviation of fluctuations of the average power. When averaging spectra calculated from a digitized data stream, $\sigma_F$ behaves as $\tau^{-1/2}$. Now, instead, we have a coherent signal that can beneficially be recorded in a single continuous stream. In the latter case, $\sigma_F$ is independent on $\tau$, but the signal builds up coherently, $P_G \propto \tau$ leading  to a much more favorable scaling.
Incidentally, \eref{eq:noiseband} can be interpreted to hold for $\sigma_F$ by the replacement $P_F \Rightarrow \sigma_F$, if one thinks that the signal stays constant. The relevant signal-to-noise ratio is then
\begin{equation}
\label{eq:snr}
\begin{split}
\m{SNR} = \frac{P_G}{\sigma_F} \simeq \frac{|d F_G|^2 }{S_f^{\m{eff}}(\omega_G)} \tau  \,.
\end{split}
\end{equation}
For details, see Appendix \ref{append}.

At this point we give list of experimental parameters used in the rest of the paper for evaluating the experimental opportunities, see Table \ref{tab:pars}. The force noise with these parameters, and with a measurement time $\tau = 1$ hour is plotted in \fref{fig:forcenoiseT} as a function of frequency around the mechanical peak. Although the mechanical noise and susceptibility have opposite effects on the sensitivity independent of frequency, the best sensitivity is reached on resonance because of the imprecision noise plays the least role there. At a better imprecision noise, the frequency dependence becomes less pronounced and the measurement hence perhaps easier. Otherwise, a near-quantum limited amplifier in this example is not important because of the overwhelming mechanical thermal noise.

We also plot in \fref{fig:forcenoiseT} the predicted gravitational signals $\sqrt{P_G}$ at different excitation amplitudes $d x_S$ of the source mass. If the effective force noise is below the gravitational signal ($\m{SNR} > 1$), the latter is observable. We see that near resonance, an amplitude $d x_S = 1$ $\mu$m exposed to our default $Q=10^7$ oscillator provides $\m{SNR} > 1$ within a very reasonable integration time. It could even be possible to reach down to $d x_S = 100$ nm if $Q \sim 10^8$ (which may not be realistic), or by integrating several days.

\subsection{Gravitational forces in a system of quantum oscillators}

As mentioned, measuring gravity from a source mass that is in a spatially separated superposition seems out of reach for quite a while, although levitated nanoparticles \cite{Aspelmeyer2020Levit,Novotny2020levit} are showing promise to this end. The first logical steps towards experimental tests of such a genuine quantum gravity would be to study gravity in a system which exhibits some macroscopic quantum properties of massive objects. The latter could include the test mass, the source mass, or both. The states under consideration are pure states that are displaced by the gravitational field. The list of foreseeable possibilities, in an expected order of experimental difficulty, could look as follows:
\begin{enumerate}[label=(\arabic*)]
    \item \label{item:gndprobe} Ground state cooling of the test mass oscillator.
    \item \label{item:gndsource} Ground state cooling of the source mass oscillator.
    \item \label{item:sqtest} Squeezing of the test mass oscillator.  
    \item \label{item:sqsource} Squeezing of the source mass oscillator.
    \item \label{item:entangle} Prepare the system of both oscillators in a two-mode squeezed state.
\end{enumerate}

In items \ref{item:gndprobe}, \ref{item:sqtest}, the test mass oscillator's state is close to a pure quantum state. In the sense that the gravity is from an equally small source mass, the situation is more quantum plus gravitational than atomic particles in Earth’s gravitational field \cite{Overhauser1975,Strelkov2002neutron,Chu1999,Rosi2014atomGR}. In case \ref{item:sqtest}, moreover, the test mass state exhibits some nonclassical properties. In item \ref{item:gndsource}, gravity is produced by a source mass whose fluctuations are localized at the vacuum level. In case of item \ref{item:sqsource}, the gravity is produced by a source whose position is in a nonclassical state. The case \ref{item:entangle}, as a genuinely entangled and nonseparable state \cite{EPR}, is the most nonclassical, and exhibits nonlocality and gravity at the same time. Ideally the actuation of the source mass does not spoil the entanglement of the fluctuations.

Below, we will review some protocols discussed in the literature for preparation of these states, and discuss the experimental challenges. We begin with dissipative protocols that create stabilized states, but require a high quality factor of the cavity. We then review alternative solutions that work optimally in the bad-cavity limit $\kappa  \gg  \omega_0$. It will be found that cooling via measurement appears the most promising method to operate our oscillators in the quantum regime given most plausible experimental parameters.

\subsubsection{Detection of gravitational signal in the quantum limit}

Before discussing the protocols to prepare the states, we will consider detection of the gravitational signal. We assume an oscillator that has been sideband-cooled close to the ground state, $n_m = 1$.  One should bear in mind that although dissipative cooling reduces the noise, the responsivity to external forces is reduced similarly. In other words, the product $\gamma n_m^T$ in Eqs.~(\ref{eq:sfefflist1},\ref{eq:sfefflist2}) stays unchanged in e.g.~sideband cooling, so that in the stationary case, force sensitivity is not enhanced (see, however discussion on non-stationary detection \cite{Woolley2020nostat}). In practical work, however, such broadening of the very sharp peak is highly beneficial, offering relaxed tolerances for driving frequencies, and stability to drift. Also, via \eref{eq:sxlistE}, the sensitivity to external vibration noise is reduced, too. 

\begin{figure}[h]
  \begin{center}
   {\includegraphics[width=0.9\columnwidth]{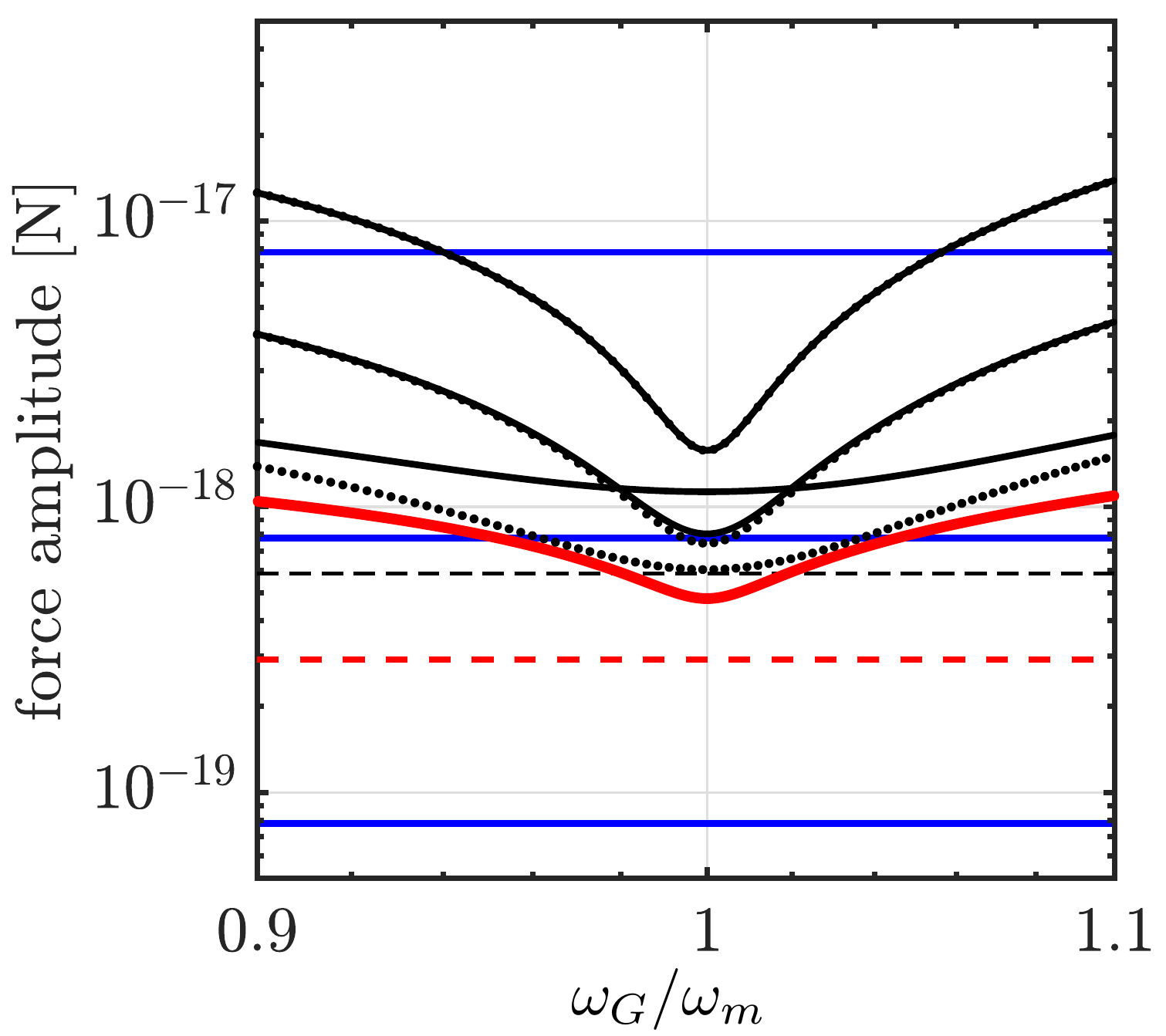} }
    \caption{\textit{Force sensitivity for a sideband-cooled oscillator.} The black curves are the effective mechanical noise at different cooperativities ($\mathcal{C} = 0.01, \, 0.1, \,  1$, from top to bottom). Additionally, the SQL is plotted with thick red line, and the dotted lines depict the results from BAE monitoring at the same cooperativities as the adjacent solid line. The horizontal lines are as in \fref{fig:forcenoiseT}. The mechanical $Q = 10
   ^7$ in all cases, and the added noise is 1 quantum. }
    \label{fig:forcenoiseQ}
 \end{center}
\end{figure}

We plot the force sensitivity for the cooled oscillator in \fref{fig:forcenoiseQ}. Because mechanical noise is nearly absent, the imprecision noise and also backaction noise start to play a role. Indeed, by comparing to \fref{fig:forcenoiseT}, it is seen that with standard (non-BAE) measurements, the signal-to-noise ratio is reduced as compared to the thermal limit. A good amplifier with $S_{\m{add}} \lesssim 1$ is needed to get within a factor of 2 from the remaining thermal noise limit, but going further than that, one needs BAE measurement. In  \fref{fig:forcenoiseQ} we also display the SQL as a red line. This is the best one can get without BAE monitoring when the oscillator is cooled to the ground state.

\subsubsection{Reservoir engineering via cavity driving}

A common protocol to prepare mechanical oscillators into motional quantum states is reservoir engineering, including sideband cooling, and single-mode \cite{ClerkSq2013} and two-mode \cite{Clerk13WangC,Meystre2013Sq,Tian13Ent,ClerkEnt2014,Vitali15Ent} squeezing. The reservoir is constructed from the optomechanical cavity, with the interaction carefully designed via a suitable pumping such that the oscillator sees a dissipative bath that pulls the oscillator into an interesting stabilized state.

A challenge in the current case is that dissipative protocols are ideally  devised for the good-cavity limit $\kappa \gg \omega_0$, which is difficult to reach with 2 kHz oscillators. Empty superconducting 3D cavities \cite{Schoelkopf2013HighQ3D}, or those with only a small chip occupation volume \cite{Schoelkopf2015CPW3D} can attain the needed range. However, with SiN microwave optomechanics, internal cavity losses have not reached below $\sim 20$ kHz \cite{Steele2015HighQ,Nakamura3D} likely due to the bulky chip assemblies that occupy a large fraction of the mode volume.

\JM{Sideband cooling and other dissipative protocols [see bad-cavity correction term in \eref{eq:sbcool}] work reasonably well if $\omega_0  \simeq \kappa$}. This ``not-bad-cavity" situation can be in reach experimentally. To begin with, we suppose $ \kappa = \omega_0  = 2\pi \cdot 2$ kHz is possible, and evaluate predictions for cooling and squeezing using our set of parameters. Verification of the mechanics' status can be done using the same techniques used earlier in the corresponding experiments, and hence we do not discuss the specifics of the detection.


We remark here that in cases \ref{item:gndprobe} - \ref{item:sqsource} listed below, the source  and test oscillators will be in their separate 3D cavities (see Section \ref{sec:cavity}), and operations on either oscillator are less likely to disturb one another in comparison to case \ref{item:entangle} where the oscillators are coupled to the same cavity mode.

%
\ref{item:gndprobe} Sideband \emph{cooling the test mass oscillator to the ground state}, $n_m \lesssim 1$, is very achievable. From \eref{eq:sbcool} we obtain $n_m \simeq 1.0$ using  $\Gamma_{\m{opt}}/2\pi \simeq 50$ Hz and a very small input microwave power $< 1$ pW. The source mass needs to be optomechanically accessible only in a minimal amount such that it can be characterized, and its vacuum gap can be large, say $h_S \sim 10$ $\mu$m, even larger if needed.

\ref{item:gndsource} \emph{Cooling the source mass oscillator to the ground state} requires the source oscillator to be well coupled to its cavity mode, i.e.,~the vacuum gap to the antennas should be as small as possible to have a decent $g_0$. At the same time, the vacuum gap $h_S$ should be large enough to avoid crashing the actuated membrane into its antenna chip. As a compromise, we choose $h_S = 2$ $\mu$m.  An additional challenge is to induce a large enough $dx_S$ now that the source oscillator has a considerably increased damping. Based on the discussion in Section \ref{sec:actuation} below, the latter is deemed possible. Ground-state cooling becomes possible at a modest $P_{\m{in}} \sim 50$ pW.

Dissipative quantum \emph{squeezing} of a mechanical oscillator is carried out using double-sideband pumping of the cavity \cite{ClerkSq2013,SchwabSqueeze,Squeeze,TeufelSqueeze}. In order to \ref{item:sqtest} \emph{prepare the test oscillator in a squeezed state}, we use the results including bad cavity corrections from Ref.~\cite{Wang2019SqCR}. Squeezing of $\sim 1.5$ dB is expected with our parameters, and at effective coupling of the red-detuned tone of $2\pi \cdot 400$ Hz, and with optimized pump power ratio. \ref{item:sqsource} \emph{Preparing the source oscillator in a squeezed state} is otherwise similar, but since $g_0$ is smaller, a higher input power $\sim 10$ pW is needed.

\ref{item:entangle} We consider the creation of \emph{two-mode squeezing}, or \emph{entanglement} of the gravitating system the most challenging technically. Dissipative protocols require at least a not-bad-cavity cavity situation, but this will be demanding in the present case where we have to couple both oscillators to the same cavity mode, while keeping the oscillators physically separated (see Section \ref{sec:cavity}). Therefore, dissipative protocols, which are well understood in standard electromechanical systems, require cavity design that currently we do not have available. Additional challenge is that the oscillators should ideally be spaced in frequency by at least $\sim 0.1 \cdot \kappa$ such that they see a different bath, which conflicts with an efficient actuation that favours equal frequencies. 

Nonetheless, we evaluate the prospects using our parameters (Table \ref{tab:pars}), using standard analysis \cite{ClerkEnt2014,Entanglement}. We find that entanglement characterized by the Duan quantity is possible for example at $\kappa/2\pi = 2$ kHz, and with the mechanical frequencies differing by 15 \%. A solution for the issue with mismatched mechanical frequencies is to start the process from a heavily damped situation (recall the dissipative protocols' performance is proportional to the decoherence rate $\Gamma$ that is not affected by e.g.~sideband cooling), which allows actuation even the frequencies are different.

%
%

\subsubsection{Ground-state cooling by measurement-based feedback}

Damping the oscillator by applying feedback built upon the measurement record, dubbed cold-damping, is a popular cooling strategy in optical cavity optomechanics \cite{Tombesi1998feedback,Cohadon1999,Poggio2007FB,LIGO2009kgQM,Kippenberg2015FB,Bowen2016sqCool,Schliesser2018FB}. To obtain fast measurements, one needs here $\kappa  \gg  \omega_0$ that is a natural parameter regime in the current case. A challenge with entering the quantum regime with feedback is that the measurements have to track the mechanics at a rate given by the rate at which a phonon is transferred with the environment. The latter, the decoherence rate, is given by $\Gamma = \gamma n_m^T$, and the measurement rate is $\Gamma_{\m{meas}} = 4 \eta G
^2/\kappa = \eta \Gamma_{\m{opt}}$. Here, $\eta$ is the detection efficiency. With microwaves, it is related to the added noise $S_{\m{add}}$. Ideal amplifier and no signal losses imply $\eta = 1$. Systems with Josephson travelling wave parametric amplifiers (JTWPA) have reached a system $\eta \sim 50$ \% \cite{Siddiqi2015Amp}, which number we use below. To reach the quantum limit, particularly to cool to the ground state, strong measurements are required: $\Gamma_{\m{meas}} > \Gamma$.

\ref{item:gndprobe}  \emph{Cooling the test mass oscillator to the ground state} is possible with the  parameters given in Table \ref{tab:pars}. We obtain \cite{Aspelmeyer2008Cool} that $n_m < 1$ is reached for example with $P_{\m{in}} = 0.1$ nW, $\kappa/2\pi = 20$ kHz, which provide $\Gamma_{\m{meas}}/2\pi \simeq 500$ Hz. At $\kappa/2\pi = 200$ kHz ground state cooling is also possible, but requires $P_{\m{in}} = 10$ nW, which otherwise is not a problem but necessitates careful canceling of the pump before the JTWPA.

\ref{item:gndsource} \emph{Cooling the source mass to the ground state}, $n_m < 1$, is feasible as well, although a smaller $g_0$ necessitates a higher $P_{\m{in}} = 10$ nW, with $\kappa/2\pi = 20$ kHz.

\ref{item:sqtest} - \ref{item:entangle} \emph{Squeezing} using feedback \cite{Vitali2000FeedbackSQ,Vitali2002FB} requires interaction that currently is not available. However, squeezing can also in principle be created by combining feedback cooling and parametric modulation \cite{Bowen2011Squ,Falferi2013feedbackSQ}. The latter can be implemented (see Section \ref{sec:actuation} below) by the same means as mechanical actuation.

\subsubsection{Pulsed techniques}

Preparation of conditional mechanical quantum states using a few very strong measurement pulses has been proposed \cite{Aspelmeyer2010Pulse,Vanner2020pulse} and realized in a classical situation \cite{Aspelmeyer2013pulsed,Muhonen2019pulse}. In a single run, the measurement projects the state of the oscillator according to the selected protocol, and the performance is highly insensitive to the thermal population. The measurement pulse should be much shorter than the mechanical period, which requires a fast cavity response, i.e.,~the bad-cavity limit $\kappa \gg \omega_0$. The important parameter in these techniques is the pertaining measurement strength given as $\chi \simeq 4 \mathcal{G}/\kappa$. 
While promising in principle, the requirement for strong measurements $\chi > 1$ is prohibitive for reaching the quantum regime in this technique. Notice that very low-frequency mechanical oscillators, such as ours, are  beneficial because the requirement of large $\kappa$ is strongly relaxed, while the reachable effective coupling $\mathcal{G}$ is less sensitive to the oscillator frequency. Let us assume $\kappa/2\pi = 20$ kHz, and the input microwave power $P_{\m{in}} = 1$ nW.

\ref{item:gndprobe} \emph{Ground-state cooling of the test mass oscillator}: The parameters yield $\mathcal{G} \sim  2 \pi \times 10$ kHz, and $\chi \sim 2$. 

\ref{item:gndsource} \emph{Ground-state cooling of the source mass oscillator}: At the same parameters, we can reach $\chi \sim 0.4$.

It seems nonetheless that obtaining $\chi > 1$ maybe challenging to reach, at least for both oscillators, due to the power handling capability of microwave cavities. In particular exceeding the threshold for the case of the source oscillator would require a very high pump photon number $n_P \sim 10^{10}$ that is well above $n_P \sim 10^7$ usually used in planar microwave optomechanics. At the same time we remark that 3D cavities accept much more power, and also the pump is on only a fraction of time such that heating is strongly suppressed. However, a combination of a very low frequency and still a decent $g_0$ allow for a $\chi$ an order of magnitude higher than what is realistically possible with drumhead oscillators, or with bare SiN membranes.

A stroboscopic measurement involves repeating the fast measurements synchronized with the mechanical period. It was introduced as a means to carry out quantum backaction-evading measurements \cite{BraginskyQND}. The scheme was discussed in detail recently in Ref.~\cite{Nunnenkamp2020strobo}. The authors obtain that the requirement for the measurement strength can be relaxed in this case. The condition for hitting the quantum regime then reads according to Ref.~\cite{Nunnenkamp2020strobo} $\chi \gtrsim \sqrt{2\pi n_m^T/Q} $, which with our parameters becomes $\chi \gtrsim 0.4$. Based on this analysis, strong enough stroboscopic measurements seem in reach. They can be used for \ref{item:gndprobe}-\ref{item:gndsource} \emph{ground-state cooling}, \ref{item:sqtest}-\ref{item:sqsource} \emph{single-mode squeezing}, as well as for \ref{item:entangle}  \emph{two-mode squeezing} of two degenerate oscillators, with properly timed pulse sequences.

As a summary of this section, reservoir engineering for operation in the quantum regime is possible, but requires experimental progress to reduce the cavity losses by an order of magnitude from the state-of-the art. With currently or reasonably likely available parameters, cooling by measurement seems the most promising method for creation of the quantum states in the very-low frequency oscillators.

\section{Engineering aspects}

\subsection{Chip assembly}

\begin{figure}[t]
  \begin{center}
   {\includegraphics[width=0.8\columnwidth]{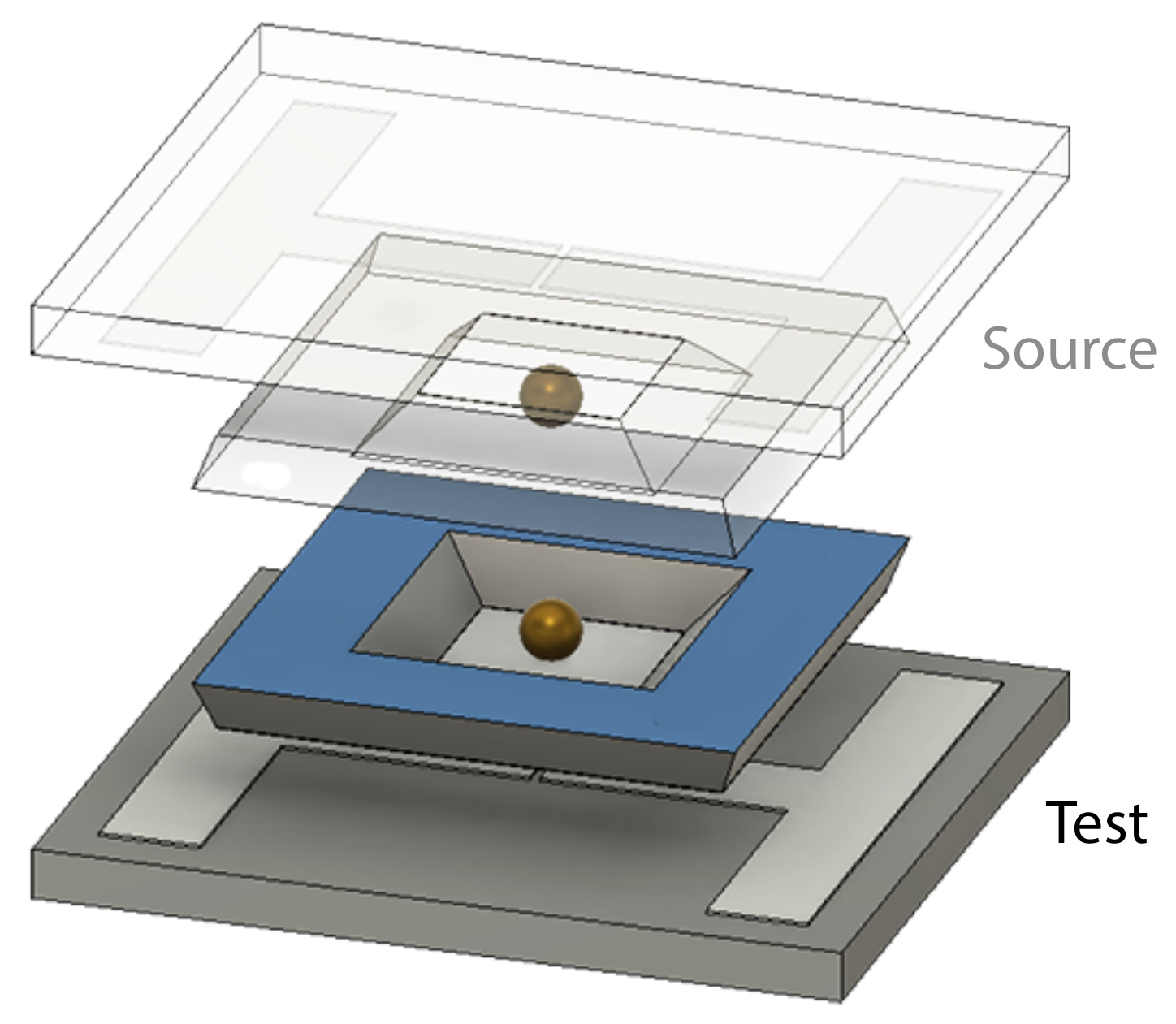} }
    \caption{\textit{Exploded view drawing of the scheme in \fref{fig:scheme1}}. In the lower set of chips (Test), the antennae for coupling to the 3D cavity are colored in white. Metallized electrodes on the membrane beneath gold are not shown.} 
    \label{fig:scheme3D}
 \end{center}
\end{figure}

Drawings of the proposed chip assemblies are shown in Figs.~\ref{fig:scheme1} and \ref{fig:scheme3D}. The test and source mass side chip assemblies are quite similar. The membrane chips are glued to their respective antenna chips. It would be tempting to proceed by rigidly attaching the two assemblies together with the gold spheres facing one another. We could not find by FEM simulations, however, a realistic way to do this without causing excessive stray mechanical coupling of the actuation into the test mass. Hence, it is better that each of the assemblies resides in their individual cavity block, which are brought close enough by nanopositioning.

\subsection{Electromagnetic design}

\subsubsection{Cavity}
\label{sec:cavity}

The design of the 3D cavity is  straightforward for the cases where individual cavities for the two oscillators can be used. However, in contrast to the chip assembly being in the center of the cavity \cite{Steele3D,Nakamura3D}, now it will be close to the wall facing the other cavity,
and the wall needs to be much thinner (clearly below 100 $\mu$m) than what typical bulky cavities exhibit. 

\begin{figure}[t]
  \begin{center}
   {\includegraphics[width=0.7\columnwidth]{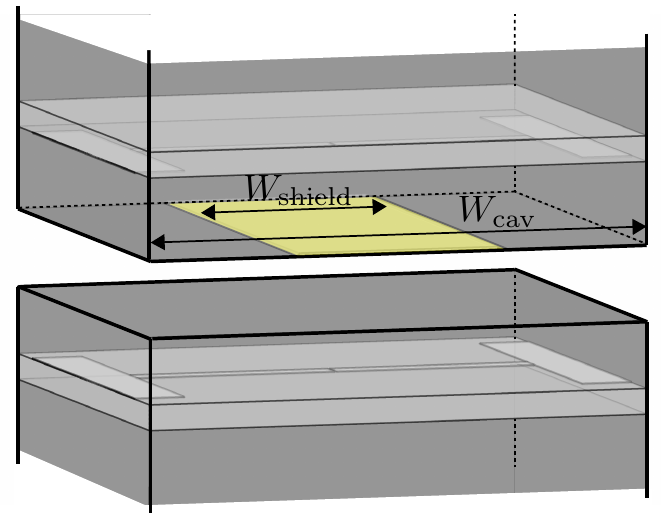} }
    \caption{\textit{Open cavity.}
    Two oscillators are coupled to the same cavity mode for creation and detection of collective mechanical states. The grounded shield (yellow) partially isolates the two blocks. Only the antenna chips are shown for clarity. The cavity volumes continue $\sim 1$ cm up and down outside the graphics. The gap between the blocks is less than a millimeter, while the chips can extend closer to one another.}
    \label{fig:cavity}
 \end{center}
\end{figure}

Coupling both oscillators to the same cavity mode, as needed for creation of two-mode squeezed states, is more involved. The cavity then consists of two blocks facing one another, with a shield in the center partially covering the opening. The design is shown in \fref{fig:cavity}. The challenge is to contain the cavity mode within the desired volume. With the complicated chip assemblies near the openings, the fields exhibit leakage that limits the cavity losses. Thus far we were not able to find a geometry where this would become insignificant given reasonable design error margins, and typically radiation losses limit the linewidth in the range $\gtrsim 100$ kHz. Thus, reaching good-cavity limit in this scheme is not readily available.

\subsubsection{Source mass actuation}
\label{sec:actuation}

The source oscillator should include frequency tuning capability in order to roughly match the mechanical frequencies. This is required for reaching a straightforward resonant actuation of the source mass, which also is felt resonantly at the test mass oscillator. The tuning can be accomplished in a standard manner by a constant voltage $V_{\m{dc}}$ applied to the antenna electrode. 
On top of the dc voltage, there is a resonant ac voltage $V_{\m{ac}}$. These two cause an electrostatic driving force amplitude
\begin{equation}
\label{eq:drive}
   F_S = V_{\m{ac}} V_{\m{dc}} \frac{d C_S}{d x }  \,,
\end{equation}
and the actuation amplitude becomes $d x_S = F_S/(\gamma \omega_0)$. Here, $C_S$ is the capacitance between the antennae and the membrane electrode. If the source oscillator is not optomechanically damped, we obtain that for example modest values $V_\m{ac} = 1$ mV, $V_\m{dc} = 0.1$ V provide $d x_S \sim 1$ $\mu$m. However, the situation changes if the source oscillator is sideband cooled close to the ground state, and for the same amplitude, we need e.g.~$V_\m{ac} \sim 2$ V, $V_\m{dc} \sim  10$ V. The latter are high values as compared to the normal range in superconducting nanoelectronics, but we still consider them manageable. Another possibility is to use a piezo shaker for actuation, instead of electrostatic actuation.


It is critical to make sure that a false positive gravitational signal is not triggered by stray coupling of the source mass actuation into the test mass oscillator. The stray coupling of mechanical actuation has to be much smaller than the corresponding gravitational signal, \eref{eq:dxT}. The ratio $d x_T/d x_S$ is typically about $-140$ dB with our parameters. One channel of stray coupling is via the electrical driving. In the suggested experiments, besides the preparation of two-mode squeezed states, the two oscillators are naturally living in their own very nearby 3D cavities, which provide an extreme isolation against stray coupling of low-frequency electric fields. Below we will present a design that describes by far the most demanding situation, that of using a single cavity mode for both oscillators, as needed for two-mode squeezing.

\begin{figure}[t]
  \begin{center}
   {\includegraphics[width=0.9\columnwidth]{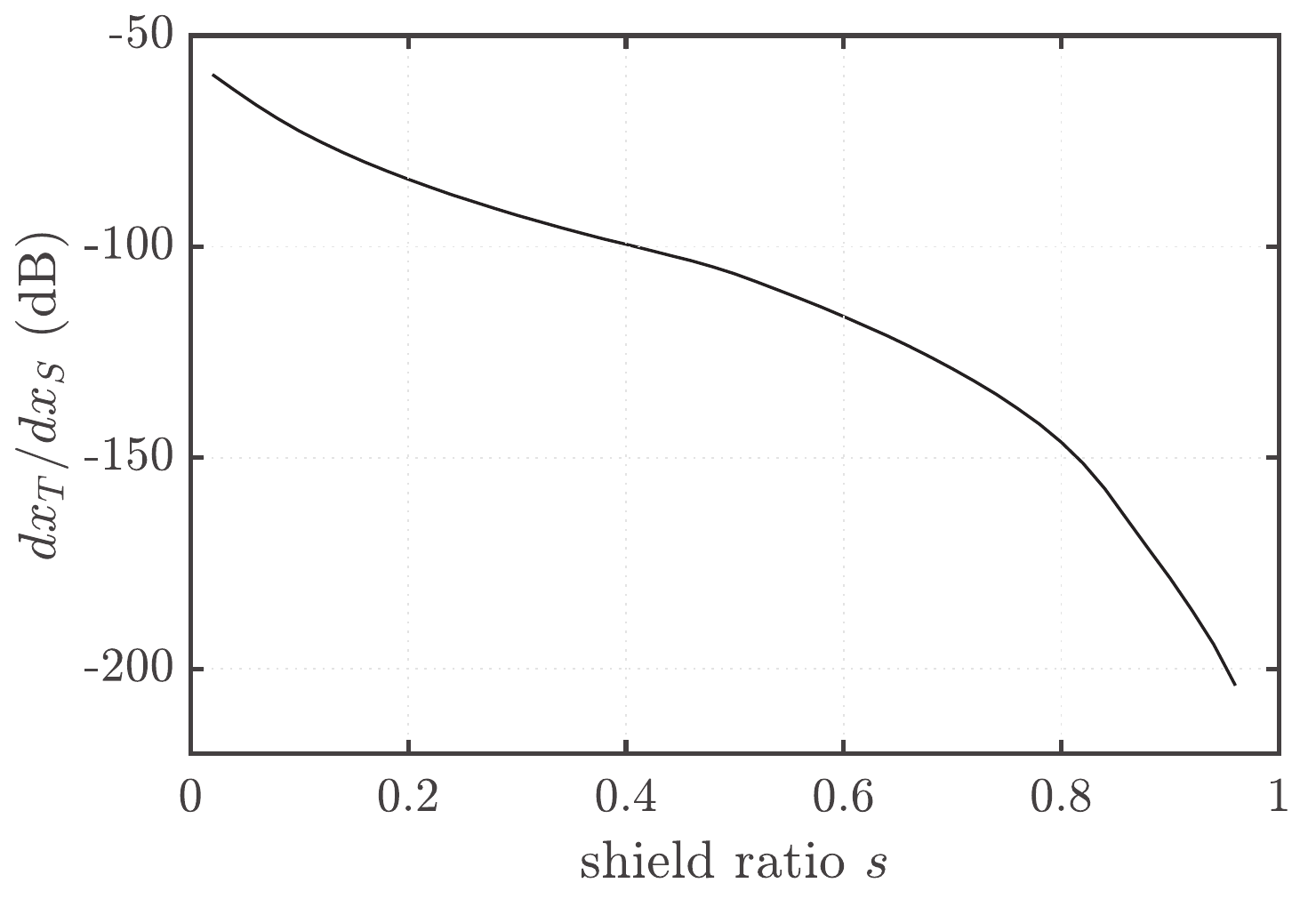} }
    \caption{\textit{Stray electrical coupling in open cavity.}
    The plot shows the stray coupling due to electrostatic actuation, as a function of a shield coverage fraction. The data is obtained from the  electric potentials at the source mass vs.~that of the test mass.}
    \label{fig:shielding}
 \end{center}
\end{figure}

The schematics of the open cavity, \fref{fig:cavity}, also shows the grounded metallic shield positioned in the middle of the cavity, between the two spheres. The shield covers a fraction $s = W_{\m{shield}}/W_{\m{cav}}$ of the opening. Here, the cavity width $W_{\m{cav}} \sim 1$ cm. Notice the antennae have to see the unshielded cavity volume in their ends in order to couple to the collective cavity mode.  From the FEM simulation, we extract the ac electric potential of the electrode beneath the gold spheres. As per \eref{eq:drive}, this gives the actuating force. We show in \fref{fig:shielding} the ratio of the resulting motion of the two spheres plus their electrode. We see that our design with $\sim 75$ \% of the cavity covered by the shield will be sufficient to attenuate the driving field by the required amount. We also find that at the room temperature conductivity of aluminium the penetration of electric fields through the shielding is significantly secondary to diffraction around the shielding.

Superconductors are expected to provide even larger shielding from electric fields.  1 $\mu$m thick superconducting shielding will attenuate high impedance near field radiation by over 400 dB (more than 100 dB better than room temperature copper at the same thickness)  \cite{Tripathi1994}. This would suggest that, as expected, our shielding will not perform worse if the shield is superconducting. We thus conclude that stray electrical coupling can be mitigated.


\subsection{Vibration isolation}

Our situation with the design of vibration isolation has some resemblance to that for gravitational wave detectors that is well established. For LIGO, the designs are very bulky and thus not directly applicable to our case that works at 2 kHz and needs to have a decently small footprint to fit into a dilution refrigerator. However, the resonant detector MiniGrail is an example of the latter, where above 300 dB of attenuation is claimed \cite{Minigrail2004}.

Isolation of our mass-loaded oscillators from vibration disturbances is critical for two reasons. There are several vibration noise sources in cryogenic systems that emit vibrations in the kHz frequency range, and can drive the modes into high effective temperatures ($n_m^T$ higher than expected based on the refrigerator temperature). Second, a mechanically coupled stay actuation from the source mass into the test mass must be mitigated. 

Luckily, the frequency range of a few kHz is favorable for the design of the isolation. The lowest pendulum modes of typical filters that involve kg range masses are clearly below our range. The wavelength is large enough such that the masses or springs act as rather ideal lumped elements, and a direct brute-force FEM simulation can be used to predict the filtering of a realistic assembly. The latter is a considerable benefit since analytical modeling that can rely on heavy approximations may not properly capture the real situation.

Let us discuss the vibration noise levels in a typical pulse-tube powered dry dilution refrigerator. Vibration noise in commercial dilution refrigerators have been reported up to 1 kHz frequencies in Ref.~\cite{Olivieri2017vibration}. They obtained $\sqrt{S_x^E}(1 \, \m{kHz}) \lesssim 10^{-13}$ m$\rthz$. From the analysis in Section \ref{sec:observegrav} (see also Figs.~\ref{fig:forcenoiseT} and \ref{fig:forcenoiseQ}), we judge a scale  $\sqrt{S_x^E}(2 \, \m{kHz}) < 10^{-19}$ m$\rthz$. These numbers indicate that a very substantial attenuation of 120 dB between the refrigerator and the sample is required. One could expect that optimizations of the refrigerator configuration will lead to lower initial noise levels. Encouragingly, in Ref.~\cite{Oosterkampc2018isolation}, 3 kHz oscillator was successfully thermalized down to $\sim 30$ mK when filtering was installed in a dry dilution refrigerator.

Similar to the case of electrical stray coupling (Section \ref{sec:actuation}), the stray coupling of mechanical actuation has to be much smaller than the gravitational signal. The requirement is around $-140$ dB with our parameters. An estimate is then that there should be at least this much attenuation between the anchoring points of the two chip assemblies. We use as a criterion the attenuation between anchoring points of the 3D cavities, which is a worst-case estimate since in reality there will be attenuation between the chip and the cavity anchoring. Our isolation requirement thus becomes: $120$ dB between either oscillator and the refrigerator, and $140$ dB between the oscillators (in the classical experiment the source oscillator does not necessarily need isolation).


\begin{figure}[t]
  \begin{center}
   {\includegraphics[width=0.95\columnwidth]{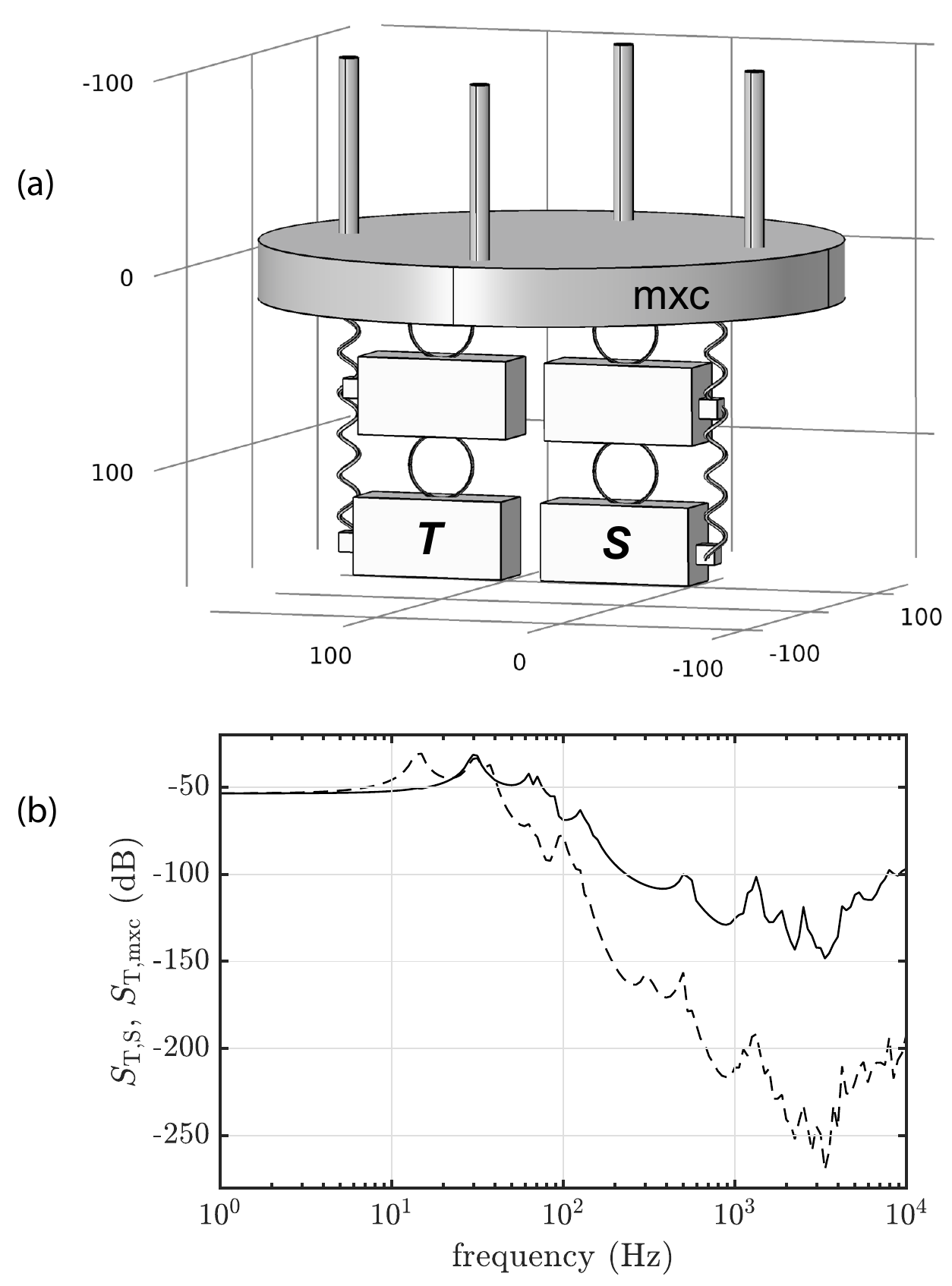} }
    \caption{\textit{Design of the suspension and vibration isolation.} (a) Cavities containing the oscillators are each attached to the bottom of their own mechanical filtering systems. The filters hang from the mixing chamber plate (mxc) of the dilution cryostat. One of the filters has cryogenic nanopositioning system (not shown). The helices model flexible thermal contacts and/or cables. The scale is in millimeters. (b) Finite-element simulation of the transfer function of displacement excitation between the mxc plate and the test mass block (solid lines), and between the test mass and source mass blocks (dashed lines).}
    \label{fig:suspension}
 \end{center}
\end{figure}

Our design is shown in \fref{fig:suspension} (a). The vertical bars at top are regular support tubes of the cryostat, made of stainless steel, which is also the material we assume for the ring springs. Otherwise, the material is copper. For the FEM simulation, we supposed a rigid anchoring from the top surfaces of the support tubes. The wide plate is the mixing chamber plate. Below this there are hanging two suspension assemblies, one for either oscillator. The assemblies consist of two springs and two masses. Coil springs are questionable due to the complicated mode structure they posses in our frequency range. Instead, we adapt ring springs \cite{Minigrail2004,Oosterkamp2014stm} that also keep the assembly more rigid and easy to control in a confined space, without compromising the filtering capability. The devices are attached to the lowermost masses, and can be moved by nanopositioning system e.g.~between the mixing chamber and the first springs. We employed a modest underdamping of the lowest modes, which can be realized by magnets. Asymmetry in the structure was included in order to model a realistic case. The weight of the masses $\sim 1$ kg is selected such that the pendulum modes are below $\sim 100$ Hz. The mixing chamber plate needs to be rather massive, since otherwise its lowest flexural modes are close to the band of interest.

Figure \ref{fig:suspension} (b) displays, first of all, the predicted attenuation between the lowermost masses of the two legs, shown in a dashed line. The peaks at $\sim 500$ Hz and $\sim 1.2$ kHz are flexural modes of the plate, and the raise above 3 kHz is due to modes in the thermalizations. The attenuation between the masses around 2 kHz becomes $\sim 220$ dB, and between the refrigerator and a single mass $\sim 120$ dB. As a result, we can conclude that a rather simple filtering setup essentially satisfies the requirements, and leaves possibilities for development.




\subsection{Phase noise of the microwave source}

Microwave optomechanics needs ultrapure sinewave sources. High-frequency phase noise at frequencies offset from the pump by the mechanical frequency has to be suppressed below the vacuum level at the sample, otherwise this noise will appear as an extra heating source. The suppression is normally done by room-temperature tunable cavity filters typically in the amount of tens of dB.

In our current case, the challenge with filtering seems at first sight drastically increased as compared to e.g.~experiments with 10 MHz mechanical devices. First of all, the phase noise grows unavoidably towards lower frequency offsets from the pump. Second, our samples have much smaller $g_0$, which seems to pose a demand for high pump powers and hence more filtering since the phase noise floor is a certain amount below the pump power. Third, one needs filters with a very narrow bandwidth, down to only a few kHz, which is not feasible with room-temperature tunable cavities.

Let us estimate the needed filtering. The input power $P_{\m{in}}$ and cavity photon number are related as
\begin{equation}
\label{eq:nc}
n_P =  \frac{P_{\m{in}} \kappa_e}{\hbar \omega_c} \frac{1}{\Delta^2 + \LL(\frac{\kappa}{2}\RR)^2}  \,,
\end{equation}
where $\Delta$ is the detuning of the pump frequency from the cavity. In the case of dissipative protocols, $|\Delta| \approx \omega_0$, and in measurement based protocols $\Delta \approx 0$. Good commercial microwave generators have phase noise specifications at 2 kHz of $\sim -110$ dBc/Hz, which means that the noise floor is by this amount below the pump level. Now we should compare the pump power at the sample plane, at a given effective coupling, with the noise floor. We assume the bad-cavity limit, $\kappa = 10 \, \omega_0$. Using the parameters of the test mass oscillator in Table \ref{tab:pars}, we obtain that in order to have a large measurement strength $ \sim 1$ kHz, the input power is $P_{\m{in}} \sim 0.1$ nW. The noise floor is then $\sim 20$ dB above the vacuum level, which requires suppression by a rather modest amount $\sim 25$ dB. The requirement is substantially relaxed if one manages to obtain a not-bad cavity situation for the dissipative protocols, and barely any filtering will be needed.

We hence obtain that the phase noise suppression requirement is barely affected in comparison to state-of-the-art microwave optomechanics. This is associated to the fact that in the bad cavity limit, \eref{eq:nc} goes as $\propto (\kappa)^{-1}$, and $\kappa$ can be set according to $\omega_0$ by the external coupling.

The still remaining issue is the need for narrow-band notch filters. They can be implemented using a tunable superconducting cryogenic cavity, which can operate at 4 Kelvin environment.

\subsection{Other forces}

There are several forces between nearby objects that can mask the gravity. The most famous is the Casimir force, which can be very strong at small separations. Its zero-temperature expression for a sphere of radius $R$ and a plate at a distance $d$, in the limit $R \gg d$, reads \cite{Lamoreaux_1997}
\begin{equation}
  F_C = - \frac{\pi^3 \hbar c R}{360 d^3} \, . 
  \label{eq:PFA_casimir}
\end{equation}
In our experiment, the Casimir force between the source mass and the shielding varies in time. This will cause the shielding to oscillate, and in turn the Casimir force on the test mass will also vary. If the distance in \eref{eq:PFA_casimir} varies sinusoidally with amplitude $d x_S$ around a mean value $d$, then the time dependent component will reduce to
\begin{equation}
  |d F_C| = \frac{\pi^3 \hbar c R}{120 d^4} d x_S \, . 
  \label{eq:Casimir_Driving}
\end{equation}
The amplitude of the oscillating force between the shielding and the test mass is strictly less than that between the source mass and the shielding, only becoming equal in the limit of the Casimir force dominating the modulus of the shielding. This allows us to calculate an absolute upper bound on the Casimir driving using \eref{eq:Casimir_Driving}. If the source mass is at a mean distance of $d=100$ $\mu$m from the shielding and is sinusoidally driven with an amplitude $d x_S = 1$ $\mu$m then the amplitude of the Casimir driving will be less than 20 zN. This is  three orders of magnitude smaller than the gravitational force for a symmetrical setup with the same dimensions.
%
%

Other forces are discussed e.g.~in Ref.~\cite{Aspelmeyer2016grav}. These include those due to free charges, or electrostatic patch potentials. We find that on the order 100 charges on the spheres at $d = 100$ $\mu$m separation gives a force equal to the gravity. Characterizing the amount of charges, and getting rid of them in a cryogenic experiment will be difficult.

\begin{figure*}[ht]
  \begin{center}
   {\includegraphics[width=0.95\textwidth]{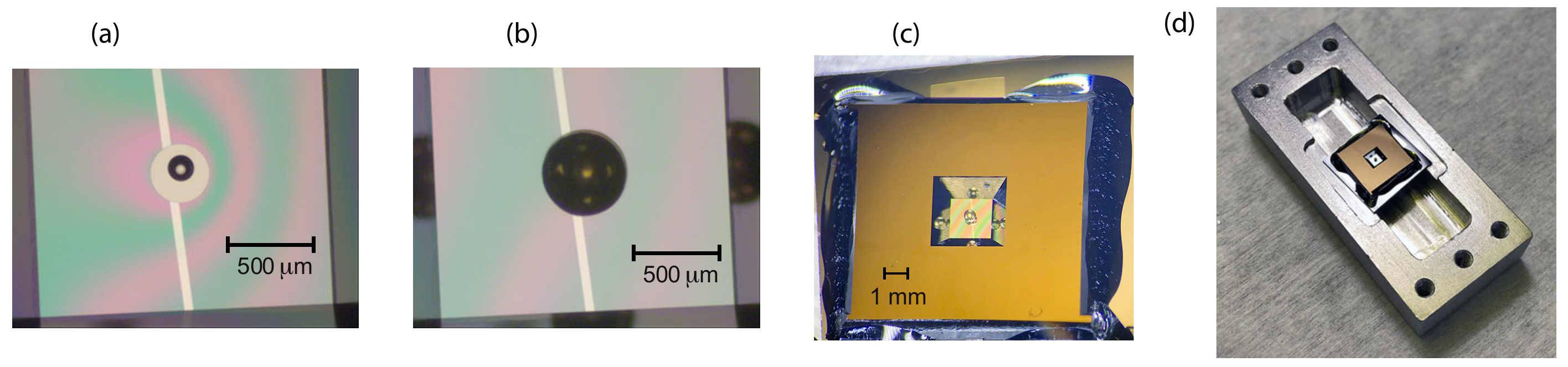} }
    \caption{\textit{Preparing the sample.} (a) The membrane chip with 1.7 mm wide, 50 nm thick SiN window has been flipped on top of an antenna chip, and firmly glued into it. Epoxy glue drop about 150 microns in diameter is micropositioned in the center of the membrane on top of an Al electrode. (b) 0.5 mm diameter gold sphere is micropositioned on top of the glue. (c) Photograph of entire chip. Reflections of the gold sphere appear at the oblique mirror-like sidewalls. (d) The assembled device attached to the lower half of the rectangular 3D microwave cavity. }
    \label{fig:glueing}
 \end{center}
\end{figure*}

In spite of the other forces, some of which are very difficult to independently calibrate out, we foresee that they do not pose a major challenge. First, we assume there is a grounded conductive shield between the spheres. Similar to the case of low-frequency electric fields (see Section \ref{sec:actuation}), we argue that even a partial shield is very effective at canceling these forces that occur between adjacent surfaces. Second, gravity has a unique signature that it depends on the COM distance, whereas the other forces depend on the surface separation. Tuning the distance between the spheres can thus facilitate spotting the gravity.

We remark that at deep cryogenic temperatures, damping by collisions with gas molecules is not an issue. We are not aware of pressure estimates inside dilution refrigerator, but one can expect that essentially all the particles will stick upon collision owing to a vanishing vapor pressure. If the experimental volume is shielded from outside, the pressure is zero for all practical purposes.

\section{Experiment with mass-loaded membrane}

We carried out a test of a fundamental part of the proposal, namely a single membrane oscillator loaded by a 0.5 mm diameter gold sphere. The goals were to verify the predicted mode structure, and to show the possibility for high mechanical $Q$ values as predicted by simulations.

The membrane window chip we discus below was purchases from Norcada. It was (1 mm x 10 mm x 10 mm) sized Si chip coated by 50 nm of SiN with roughly 900 MPa tensile prestress. The SiN window was a 1.7 mm wide rectangle made using KOH etch. Aluminum metallization was applied in the center of the membrane as a 200 $\mu$m circle. The antenna chips, of standard FZ silicon, were Al-metallized as well. The membrane chip was flipped on top of the antenna chip. The natural gap between the chips without any spacers in between becomes typically less than a micron, set by dust or irregularities, although in this assembly we believe the gap was on the order a few microns. After flipping, the membrane chip was glued with epoxy from one corner, and after the epoxy had cured, firmly glued around all edges. The latter was because we expect there can be modes involving the entire chip in the kHz range, if the chip is not fully anchored. The 4N gold spheres were purchased from Goodfellow. The fabrication steps are shown in \fref{fig:glueing} for a device similar to the one measured. A sphere was positioned on the membrane as shown in \fref{fig:glueing} (a-c). The chip assembly was fixed using PMMA resist into a standard 3D copper cavity [\fref{fig:glueing} (d)].

The measurements were carried out in a dry pulse-tube dilution refrigerator at around 10 mK temperature. The sample stage had a copper braid suspension, but there was no other vibration damping in the setup. We found that normal operation of the refrigerator when the pulse tube is running caused the mechanical modes to be strongly excited. Thus the data were recorded when the pulse tube was switched off momentarily.

\begin{figure*}[t]
  \begin{center}
   {\includegraphics[width=0.95\textwidth]{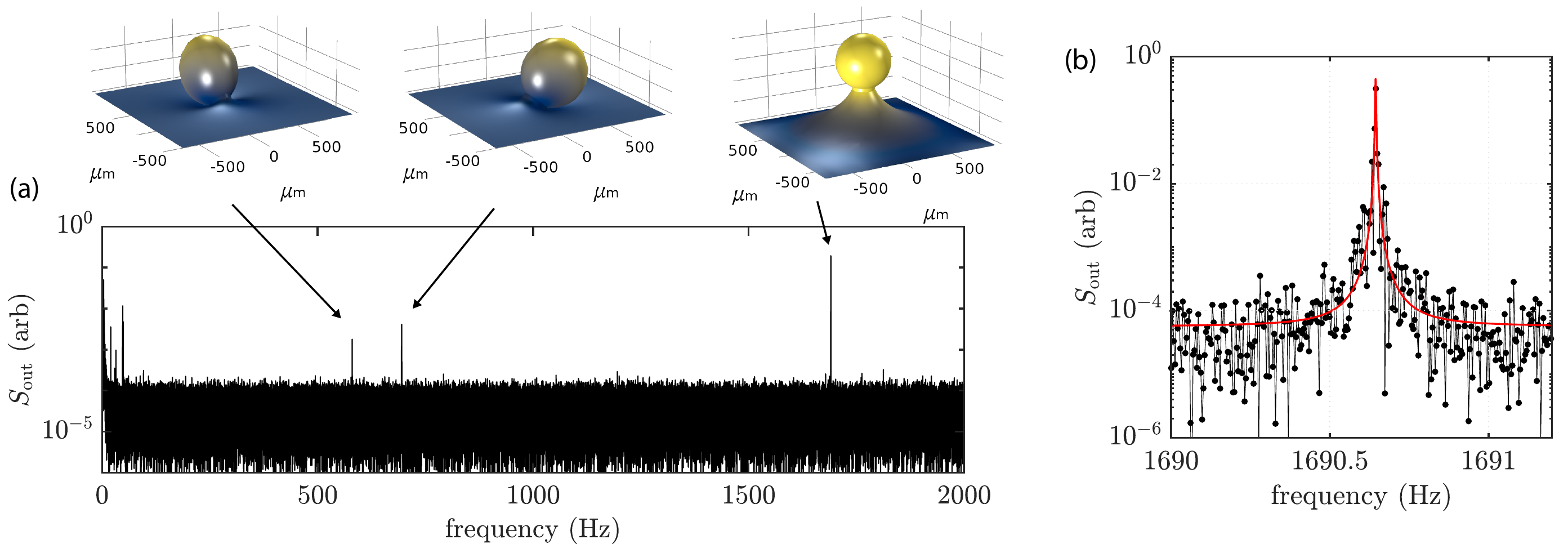} }
    \caption{\textit{Mechanical noise spectrum at 10 mK.} (a) Large frequency range with respect to the pump frequency. Three peaks interpreted as drum modes are visible. The corresponding mode profiles predicted by finite-element simulation are displayed. (b) Zoom into the lowest drum mode with $f_0 \simeq 1.69$ kHz, and $Q \simeq 6.4 \times 10^5$. The red line is a Lorentzian fit.}
    \label{fig:expdata}
 \end{center}
\end{figure*}

In \fref{fig:expdata}(a) we display the measured mechanical noise spectrum over a large frequency range. The strongest peak at $\omega_0/2\pi \simeq 1.69$ kHz is identified as the desired flexural mode. The peaks below 100 Hz are likely motion of the entire chip assembly. The two smaller peaks at 580 Hz and 690 Hz are interpreted as the two wobbling modes, which naturally have much lower optomechanical coupling. The mode profiles are also displayed in \fref{fig:expdata}(a). The measured $\omega_0$ agrees well with the result from \eref{eq:f0load} and the FEM simulation with the known parameters and the glue radius $r_{\m{glue}} = 100$ $\mu$m. The wobbling modes in a fully symmetric case are degenerate, but this is lifted by asymmetries in e.g.~the positioning of the glue and the sphere. However, by distorting the symmetry in simulations by a reasonable amount, we were not able to reproduce such a high nondegeneracy. The reason could be distortion of the chip due to thermal contraction during cooldown, which currently is beyond scope of the simulations.


The mechanical peak of the main mode is displayed in \fref{fig:expdata}(b), where we plot the total spectrum $S_{\m{out}} (\omega) = S_{\m{x,out}} (\omega) + S_{\m{y,out}} (\omega)$. The $Q$ value extracted from the fit is $Q \simeq 6.4 \times 10^5$. The $Q$ can be even higher, since the frequency resolution was limited by the time $\sim 5$ min we were able to run the fridge without the pulse tube running. Supposing the $Q$ is limited by losses in  the adhesive, its loss factor in the idealized geometry becomes $\eta_{\m{glue}} \sim 2 \times 10^{-2}$ (see Table \ref{tab:simQ}). Using e.g.~PMMA \cite{Thompson2002materials} ($\eta \sim 10^{-3}$) as the glue, one could further improve the $Q$ value. We also expect the details of the glue attachment point can be optimized.


The cavity frequency with the chip inside was 5.27 GHz, and internal linewidth 9 MHz. The latter is clearly higher than $\approx 0.3$ MHz we usually observe with our standard bare 0.5 mm wide membrane chips in similar (non-superconducting) cavities. We do not attribute the high cavity losses to the gold spheres or the assembly in general, nonetheless. With a similar bare 1.7 mm window chip, the cavity losses are also significantly enhanced. Moreover, a standard 0.5 mm window chip loaded by the gold sphere, measured in an aluminum cavity, resulted in 0.3 MHz internal linewidth. Based on the measured cavity properties, one can see, however, that reaching the not-bad-cavity case with a few kHz oscillators is not straightforward.

\section{Conclusions}

In conclusion, we have presented a detailed experimental proposal for; (1) Observing gravity between milligram range masses using microwave optomechanics at deep cryogenic temperatures; (2) Pushing the experiment down to the quantum limit where the positions of source mass or test mass, or both, exhibit significant quantum fluctuations. Although the massive quantum state will be created microwave-optomechanically and not through gravity, the measurement will open a new regime of experimental study on the overlap of quantum mechanics and gravity. One can even foresee that when these two disparate entities are forced closer to another than ever before, one can experimentally achieve a limit where fundamentally new physics may be revealed, possibly revising our understanding of quantum mechanics and gravity. Additionally, observing highly massive mechanical oscillators in the quantum regime is, in itself, an important step in this direction.

\begin{acknowledgments} We would like to thank Matt Woolley and Laure Mercier de L\'epinay for useful discussions. This work was supported by the Academy of Finland (contracts 308290, 307757), by the European Research Council (contract 615755), by the Centre for Quantum Engineering at Aalto University, by The Finnish Foundation for Technology Promotion, and by the Wihuri Foundation. We acknowledge funding from the European Union's Horizon 2020 research and innovation program under grant agreement No.~732894 (FETPRO HOT). We acknowledge the facilities and technical support of Otaniemi research infrastructure for Micro and Nanotechnologies (OtaNano).
\end{acknowledgments}

\appendix
\label{append}

\section{Signal-to-noise in the detection of a coherent signal}

We discuss here an important issue that is often overlooked. The question is simple: How long does one have to average a coherent signal in order to see it?
The derivation below is not fully rigorous, but we hope it will clarify the issue. In order to properly analyze the situation, one has to know how the measurements are performed and processed inside an instrument (such as spectrum analyzer). Since this information is not easy to find or digest, we select a conceptually simple but even the most important measurement system: data digitizer. It can be assumed to make instantaneous snapshots of the data stream.

We discuss the quantity $x(t)$, such as the amplified voltage from the sample. It is characterized by the correlation function $R(t)$ and the spectral density $S(\omega)$:
\begin{equation}
\begin{split}
& R(t)= \langle x(t_0) x(t_0-t) \rangle_{t_0} = \frac{1}{T_0} \int^{T_0/2}_{-T_0/2} dt_0  x(t_0) x(t_0-t) \\
& S(\omega) =  \mathcal{F}[R(t)]
= \int d \omega R(t) \exp(-i\omega t) \,.
\end{split}
\end{equation}

We suppose the measurement of $x(t)$ is a continuous sampling that lasts over the time $\tau$. Let us sample the quantity $x(t)$ at time intervals $T$. Then there are a total of $N = \tau/T$ data readings. The sampling provides the readings $x_n = x(nT)$, where $n$ goes from 0 to $N-1$. 

In order to satisfy the Nyquist sampling criterion, before sampling the data has to be restricted into the spectral bandwidth
\begin{equation}
\begin{split}
\label{eq:band}
|\omega_B/2\pi| = |f_B| < (2T)^{-1}
\end{split}
\end{equation}
by analog low-pass filtering. The frequency response of an ideal low-pass filter equals the box function; $U(\omega) = 1$ when $-\omega_B < \omega < \omega_B$, and $U(\omega) = 0$ elsewhere. In time domain, the filtering function is
\begin{equation}
\begin{split}
 u(t) & = \mathcal{F}^{-1} [U(\omega)] = \frac{1}{2\pi} \int d \omega U(\omega) \exp(i\omega t) \\
 & = \frac{\omega_B}{2\pi }   \m{sinc}(\omega_B t)  \,.
\end{split}
\end{equation}
After filtering, the spectral density and the correlation function are
\begin{equation}
\begin{split}
\label{eq:SBRB}
S_B(\omega) & = U(\omega) S(\omega) \,, \\
 R_B(t)& =  \mathcal{F}^{-1}[S_B(\omega)] \,.
\end{split}
\end{equation}

Below we will use the following discrete correlator of the sampled data:
\begin{equation}
\begin{split}
\label{eq:Rnm}
& R_{nm}  \equiv \langle x(nT) x(mT) \rangle \\
& = \int dt' \delta(t'-nT)   \int dt'' \delta(t''-mT) \langle x(t') x(t'') \rangle \\
& = R_B(T(n-m))
\end{split}
\end{equation}
For the last form in \eref{eq:Rnm}, we used time translation invariance.

Aiming at obtaining the spectral density, the processing starts by evaluating the discrete Fourier-transform:
\begin{equation}
\begin{split}
X_k = \sum_{n=0}^{N-1} x_n \exp(-2i\pi nk /N) \,.
\end{split}
\end{equation}
Here, the index $k$ refers to the frequency values. An estimate of the power spectral density $S(\omega)$ is then given by
\begin{equation}
\begin{split}
\label{eq:periodog}
S_k = \frac{1}{N} |X_k|^2 = \frac{1}{N} \sum_{n,m} x_n x_m \exp[2i\pi(m- n)k/N ] \,.
\end{split}
\end{equation}
Equation (\ref{eq:periodog}) is known as the periodogram. The interesting figures now are the average and variance of the periodogram. The average is
\begin{equation}
\begin{split}
\label{eq:Skavg}
\langle S_k  \rangle = & \frac{1}{N} \sum_{n,m} R_{nm} \exp[2i\pi(m- n)k/N ] \,.
\end{split}
\end{equation}
For evaluating the variance
\begin{equation}
\begin{split}
\m{Var}[S_k] = \langle S_k^2 \rangle -  \langle S_k \rangle^2  \end{split}
\end{equation}
we need the second moment of the periodogram:
\begin{equation}
\begin{split}
\label{eq:Sk2}
 \langle S_k^2 \rangle =& \frac{1}{N^2}\sum_{n,m,p,q} \langle x_n x_m x_p x_q \rangle \exp[2i\pi(m- n + p-q)k/N ]
\end{split}
\end{equation}

\subsection{White noise}

We now treat the case that the incoming signal is uncorrelated white noise: $R(t) = S_0 \delta(t)$, and $S(\omega) = S_0$. The filtered correlation function in \eref{eq:SBRB} becomes $R_B(t) = S_0 u(t)$, with the discrete version [\eref{eq:Rnm}]
\begin{equation}
\begin{split}
\label{eq:Rnmwhite}
R_{nm} = S_0 \frac{\omega_B}{2\pi }   \m{sinc}[\omega_B T(n-m)] 
= S_0 \frac{\omega_B}{2\pi } \m{sinc}[\pi (n-m)] \,.
\end{split}
\end{equation}
For the latter form, we used \eref{eq:band}.
Equation (\ref{eq:Rnmwhite}) is used in \eref{eq:Skavg} to evaluate the average spectrum. The $\m{sinc}(y)$ function can be approximated to be zero at argument values $y \gtrsim 1$. Therefore, only $n=m$ gives a contribution to $R_{nm}$. The complex exponential in \eref{eq:Skavg} in principle makes the terms oscillate as a function of $n$ and $m$. However, the condition $n=m$ also guarantees the exponential factor equals one. The average value of the sampled white noise then becomes
\begin{equation}
\begin{split}
\label{eq:Skavgresult}
\langle S_k  \rangle = S_0 f_B \,,
\end{split}
\end{equation}
which is the anticipated result of noise power within the bandwidth defined by the analog low-pass filter. Notice the result does not depend on the sampling rate nor the averaging time.

The fourth moment $\langle x_n x_m x_p x_q \rangle$ of the signal in \eref{eq:Sk2} can be written, using Wick's theorem, in terms of second moments:
\begin{equation}
\begin{split}
\label{eq:wick}
& \langle x_n x_m x_p x_q \rangle= R_{nm} R_{pq} + R_{np} R_{mq} + R_{nq} R_{mp} \,,
\end{split}
\end{equation}
where $R_{ij}$ are those given in \eref{eq:Rnmwhite}. Again, in order to obtain non-zero contributions, we require $i=j$ in all $R_{ij}$ in \eref{eq:wick}. The two first terms on the rhs are treated similar to the discussion following \eref{eq:Rnmwhite}. For the last term $R_{nq} R_{mp}$, the requirement reads as usual $n=q$, $m=p$, however the exponent in \eref{eq:Sk2} cannot equal to zero simultaneous to this requirement. Therefore, this term gives oscillatory values whose contribution will vanish. Equation (\ref{eq:Sk2}) becomes
\begin{equation}
\begin{split}
\label{eq:Sk2result}
 \langle S_k^2 \rangle & =  2 S_0^2 f_B^2 \,, \\
\end{split}
\end{equation}
and the standard deviation is
\begin{equation}
\begin{split}
\label{eq:std}
\m{std}[S_k] & =  S_0 f_B \,.
\end{split}
\end{equation}
Equation (\ref{eq:std}) is the main quantity of interest. It describes the ``noise of noise", fluctuations of the spectrum around the mean value around \eref{eq:Skavg}. If a signal peak in the spectrum is larger than $\m{std}[S_k]$, the signal becomes visible (in practice, the signal peak should be a small numerical factor larger than  $\m{std}[S_k]$).

Notice that $\m{std}[S_k]$ does \emph{not} depend on the integration time. This seems at first sight to be in stark contrast to intuition telling that uncertainty does decrease with longer integration time. In order to resolve this controversy, we have to make more detailed considerations of the measurement.

\subsection{Coherent signal}

The main model of detection of an external force is via coherent actuation of the oscillator. This is also the case in the current work. We will now calculate the average spectrum, \eref{eq:Skavg}, for a sinusoidal signal
\begin{equation}
\begin{split}
\label{eq:sinemodel}
F(t) = dF \sin(\omega_G t) \,.
\end{split}
\end{equation}
Notice that in the classical situation, its variance is zero because the signal is deterministic. The continuous and the discrete correlation functions are 
\begin{equation}
\begin{split}
R(t) & = \frac{dF^2}{2} \cos(\omega_G t) \,, \\
R_{nm} & = \frac{dF^2}{2} \cos\left[\omega_G T (n -m) \right]  \,.
\end{split}
\end{equation}
Since the spectrum is empty anywhere else other than at $\omega_G$, the only requirement for the sampling rate is $f_G < 1/(2T)$. Moreover, the signal should pass the low-pass filter, that is, $f_G < f_B$. Given the Nyquist criterion is satisfied, the overall criterion is $f_G < 1/(2T)$.

From \eref{eq:Skavg} we see that non-oscillatory terms will be obtained only if $k = \omega_G \tau /2\pi \equiv k_G$, i.e., at the frequency of the sinewave. At this frequency we have
\begin{equation}
\begin{split}
\label{eq:sksineresult}
\langle S_{k_G}  \rangle  &= \frac{dF^2}{4 N} \sum_{n,m} [ 1 + \cos(2 \omega_G T (n -m))) + \\
& i \sin(2 \omega_G T (n -m)) ]  \Longrightarrow \frac{dF^2}{4} N = \frac{dF^2}{4} \frac{\tau}{T}  \,.
\end{split}
\end{equation}
Based on \eref{eq:sksineresult}, the power of the coherent spectral peak grows, first of all, with the integration time $\tau$. This can be interpreted so that a long measurement implies higher frequency resolution, which allows for a more accurate construction of the delta function peak. Since the noise of noise $\m{std}[S_k]$ stays constant, the signal-to-noise ratio will improve over a longer measurement as it should. 

We also see that the peak grows without limit when the sampling rate $T^{-1}$ is increased. This again seems to contradict intuition and give unlimited SNR by just sampling faster, since $\m{std}[S_k]$ does not depend on $T$. However, by sampling too fast the Nyquist criterion fails to hold, and as a result, the noise grows due to aliasing, and SNR is not improved. The optimum is reached at the threshold $f_B = 1/(2T)$, and \eref{eq:sksineresult} becomes
\begin{equation}
\begin{split}
\label{eq:sksineresult2}
\langle S_{k_G}  \rangle  &=  
P_G \tau f_B  \,,
\end{split}
\end{equation}
where $P_G = \frac{dF^2}{2}$ is the signal power.

After a surprisingly tedious calculation, we obtain the final result
\begin{equation}
\begin{split}
\label{eq:snrfft}
\m{SNR} = \frac{\langle S_{k_G}  \rangle}{\m{std}[S_{k}] } = \frac{P_G}{S_0} \tau  \,,
\end{split}
\end{equation}
which is essentially \eref{eq:snr}.
Equation (\ref{eq:snr}) is often wrongly justified by comparing the mean noise level, \eref{eq:Skavgresult}, to the signal.

The integration time needed to make the signal visible, $\m{SNR} > 1$, based on \eref{eq:snr} scales rather favorably with the noise level. In order to reach this, the detection must occur as a \emph{single time-domain run} (or built based on trigger events). In this case, the signal adds up coherently. If one averages individual periodograms instead of averaging one long periodogram, the scaling will be $\tau \propto (P_G/S_0)^2$.

We will still briefly comment on the detection of a signal peak of a finite width. This occurs when the signal is of a natural origin instead of a coherent driving signal. In cavity optomechanics, this measurement is by far more important than that using a coherent drive. Now, there is no benefit of averaging in time domain instead of averaging periodograms. The average noise level \eref{eq:Skavg} reflects the original spectral lineshape, and the noise of noise, \eref{eq:Sk2result}, does not depend on the integration time. Where does it then appear that a longer measurement should improve the SNR? A longer measurement implies more information because it increases the frequency resolution $\propto \tau^{-1}$, that is, there are more independent data points available. Adjacent frequency bins can then be processed by a moving average filter with a bandwidth less than the peak width. Such averaging of data points will then reduce the noise of noise.




%

\end{document}